\documentclass[pre,superscriptaddress,12pt,onecolumn,amsmath,showpacs]{revtex4}

\usepackage{amsmath}
\usepackage{amssymb}
\usepackage{graphicx}
\usepackage{booktabs}

\newcommand{\br}{\mathbf r}
\newcommand{\avg}[1]{\left\langle #1 \right\rangle}
\begin{document}
\title{Hanbury Brown and Twiss interferometry at a free-electron laser}
\author{A. Singer}
\author{U. Lorenz}
	\affiliation{Deutsches Elektronen-Synchrotron DESY, Notkestrasse 85, D-22607 Hamburg, Germany}
\author{F. Sorgenfrei}
	\altaffiliation{present address: Helmholtz Zentrum Berlin f\"ur Materialien und Energie GmbH, Albert-Einstein-Str 15, D-12489 Berlin, Germany}
	\affiliation{Institut f\"ur Experimentalphysik and CFEL, University of Hamburg, Luruper Chaussee 149, D-22603 Hamburg, Germany}
\author{N. Gerasimova}
\author{J. Gulden}
	\altaffiliation{present address: Institut f\"ur Regenerative EnergieSysteme, Zur Schwedenschanze 15, D-18435 Stralsund, Germany}
\author{O. M. Yefanov}
	\altaffiliation{present address: Center for Free-Electron Lasers, Notkestrasse 85, D-22607 Hamburg, Germany}
\author{R. P. Kurta}
\author{A. Shabalin}
\author{R. Dronyak}
\author{R. Treusch}
\author{V. Kocharyan}
\author{E. Weckert}
	\affiliation{Deutsches Elektronen-Synchrotron DESY, Notkestrasse 85, D-22607 Hamburg, Germany}
\author{W. Wurth}
	\altaffiliation[Corresponding author:]{Wilfried.Wurth@desy.de}
	\affiliation{Institut f\"ur Experimentalphysik and CFEL, University of Hamburg, Luruper Chaussee 149, D-22603 Hamburg, Germany}
\author{I. A. Vartanyants}
	\altaffiliation[Corresponding author:]{Ivan.Vartaniants@desy.de}
	\affiliation{Deutsches Elektronen-Synchrotron DESY, Notkestrasse 85, D-22607 Hamburg, Germany}
	\affiliation{National Research Nuclear University, ''MEPhI'', 115409 Moscow, Russia}

\begin{abstract}
We present measurements of second- and higher-order intensity correlation functions (so-called Hanbury Brown and Twiss experiment) performed at the free-electron laser (FEL) FLASH in the non-linear regime of its operation.
We demonstrate the high transverse coherence properties of the FEL beam with a degree of transverse coherence of about 80\% and degeneracy parameter of the order $10^9$ that makes it similar to laser sources.
Intensity correlation measurements in spatial and frequency domain gave an estimate of the FEL average pulse duration of 50 fs.
Our measurements of the higher-order correlation functions indicate that FEL radiation obeys Gaussian statistics, which is characteristic to chaotic sources.
\end{abstract}
\maketitle

Hanbury Brown and Twiss in their pioneering experiments \cite{1,2} demonstrated that one can get fundamental information on the statistics of light sources by measuring intensity correlations at two separated spatial positions.
Originally designed as a robust method to determine the size of stars, these experiments initiated developments in the field of quantum optics \cite{3}.
Statistical properties of thermal sources \cite{1,2}, lasers \cite{4}, semiconductor microcavities \cite{5}, and, recently, Bose-Einstein condensates \cite{6,7} have been studied using this technique.
The recent advent of x-ray free-electron lasers (FELs) \cite{8,9,10,11} with their unprecedented peak brilliance and ultrashort pulse duration has opened the route to a number of spectacular ground-breaking experiments including femtosecond nanocrystallography \cite{12} and single particle coherent imaging \cite{12a,13}.
Many of these experiments exploit the high degree of coherence of the FELs.

Coherence is the defining feature of a laser source and is described by correlation functions within statistical optics \cite{14,14a}.
The first-order correlation properties of FEL sources have been extensively investigated recently \cite{15,16,17,18,19,20}.
It was experimentally demonstrated that FELs based on the self-amplified spontaneous emission (SASE) process have a high degree of transverse coherence but poor temporal coherence.
To get a more detailed picture of the statistical properties of these sources, higher-order field correlations must be studied.
These can be explored, for example, by utilizing intensity correlation measurements in a Hanbury Brown and Twiss (HBT) experiment.
In this letter we present measurements of second- and higher-order intensity correlation functions at the free-electron laser FLASH \cite{8}.

The core idea of the HBT experiment \cite{1,2} is to determine the normalized second-order intensity correlation function
\begin{equation}
g^{(2)} (\br_1,\br_2)=\frac{\avg{I(\br_1 )\cdot I(\br_2 )}}{\avg{I(\br_1 )}\avg{I(\br_2)}},
\label{eq:1}
\end{equation}
by measuring the coincident response of two detectors at separated positions $\br_1$ and $\br_2$ (see for review \cite{21}).
In Eq.~\eqref{eq:1}, $I(\br_1)$, $I(\br_2)$ are the intensities of the wavefield, and the averaging is done over a large ensemble of different realizations of the wavefield.
It is well established that chaotic light can be described in the frame of Gaussian statistics \cite{14} and is completely determined by the first-order correlation function in spatial domain known as the normalized spectral degree of coherence (SDC) $\mu(\br_1,\br_2)$.
It is defined as \cite{14} $\mu(\br_1,\br_2)=W(\br_1,\br_2)/\sqrt{S(\br_1 )S(\br_2)}$, where $W(\br_1,\br_2)$ is the cross spectral density function and $S(\br)$ is the spectral density.

The intensity correlation function then reduces to (see the Appendix A for details)
\begin{equation}
	g^{(2)}(\br_1,\br_2 ) = 1+\zeta_2 (D_\omega)\left\vert\mu(\br_1,\br_2)\right\vert^2,
\label{eq:2}
\end{equation}
where $\zeta_2(D_\omega)$ is the contrast function that strongly depends on the bandwidth $D_\omega$ of the radiation.
It was earlier demonstrated \cite{23} that for stationary chaotic sources the contrast $\zeta_2(D_\omega)$ is determined by the ratio $\tau_c/T$, where $\tau_c=2\pi/D_\omega$ is the coherence time of the wavefield and $T$ is the time resolution of the detectors.
Equivalently, the contrast $\zeta_2(D_\omega)$ determines the number of longitudinal modes $M_T=1/\zeta_2(D_\omega)$ for a chaotic source.
By definition the SDC $\left\vert\mu(\br_1,\br_2)\right\vert\le1$ and intensity correlation function $g^{(2)} (\br_1,\br_2) \le 2$.

For the full statistical description of the wavefield the n-th order correlation functions can be introduced \cite{3,14}
\begin{equation}
	g^{(n)} (\br_1,\ldots,\br_n )=\frac{\avg{\prod_{i=1}^{n}I(\br_i)}}{\prod_{i=1}^{n} \avg{I(\br_i)} }.
\label{eq:3}
\end{equation}
For chaotic light $g^{(n)} (\br_1,\ldots,\br_n)$ is completely described by the SDC $\mu(\br_1,\br_2 )$ due to the Gaussian moment theorem \cite{14}.
A comparison between correlation functions of different orders determines whether the field obeys Gaussian statistics or not.
In particular, for chaotic light the n-th order correlation function $g^{(n)}(\br,\br,\ldots,\br)$ is equal to $n!$ for a single longitudinal mode \cite{14}.

For pulsed sources the intensity correlation measurements are naturally gated by the pulse duration T \cite{23}.
FEL sources, with pulses of few tens of femtoseconds, are ideally suited for intensity correlation measurements.
According to FEL theory \cite{24} these sources should obey Gaussian statistics in the linear and deep non-linear regime of operation.
In these conditions the correlation function $g^{(2)}(\br_1,\br_2 )$ has the form of Eq.~\eqref{eq:2} and provides access to the transverse coherence properties of an FEL as well as to its pulse duration \cite{23}.

\begin{figure}[t]
\centering
\includegraphics[width=1\linewidth]{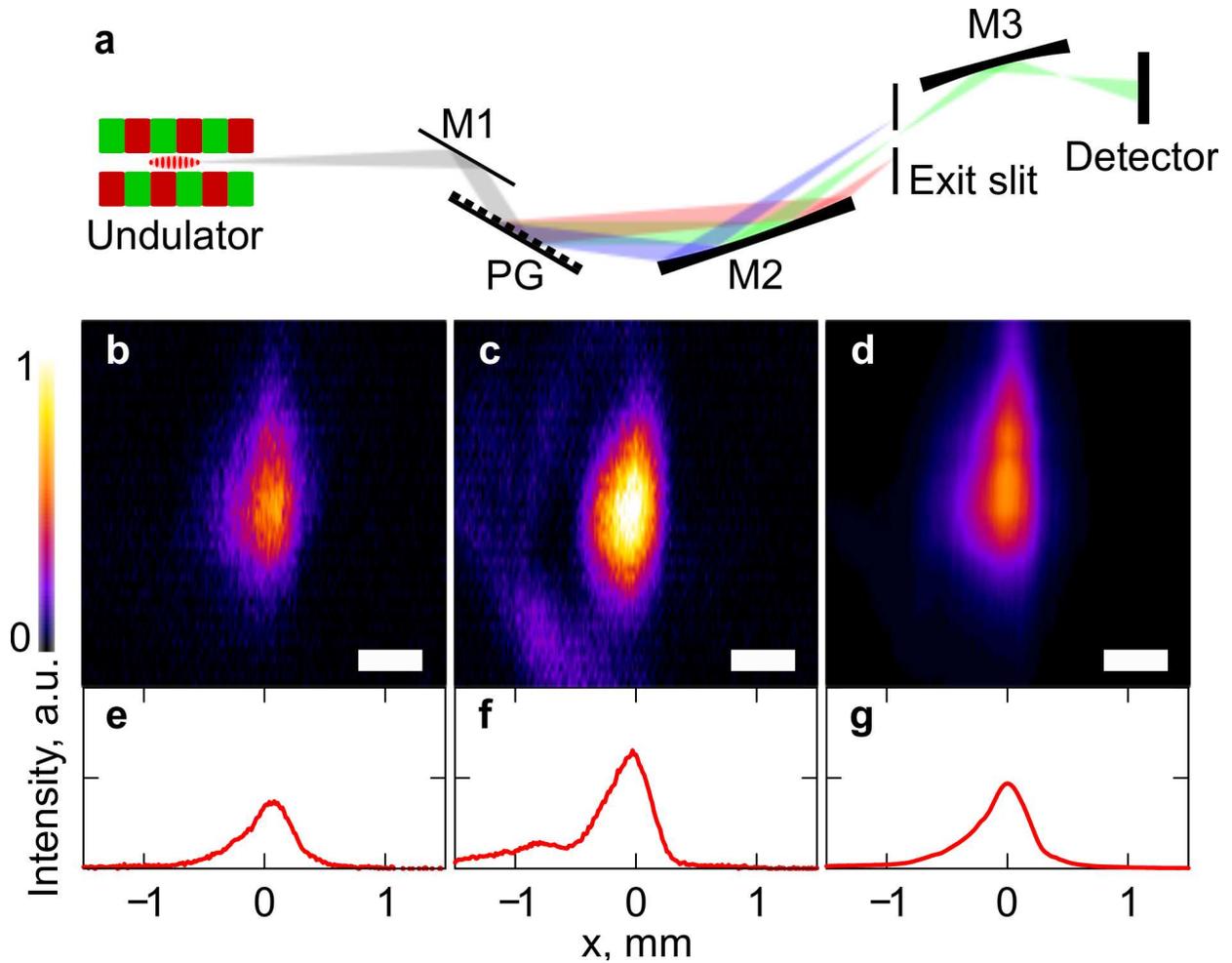}
\caption{(Color online) (a) Scheme of the experiment.
	FEL radiation is generated in the undulator and is transmitted through the beamline including three mirrors (M1, M2, M3), the plane grating (PG), and exit slit.
	Intensity profiles of individual femtosecond pulses are measured at the detector.
	(b,c) Typical single pulse intensity profiles and (d) an average over $2\cdot 10^4$ pulses for a bandwidth of $\Delta E/E=0.8\cdot10^{-4}$.
	(e-f) Projections of the pulse intensities along the vertical direction.
	The scale bar is 0.5 mm long. }
\label{fig:1}
\end{figure}
%
%
\begin{figure}[tbh]
\centering
\includegraphics[width=1\linewidth]{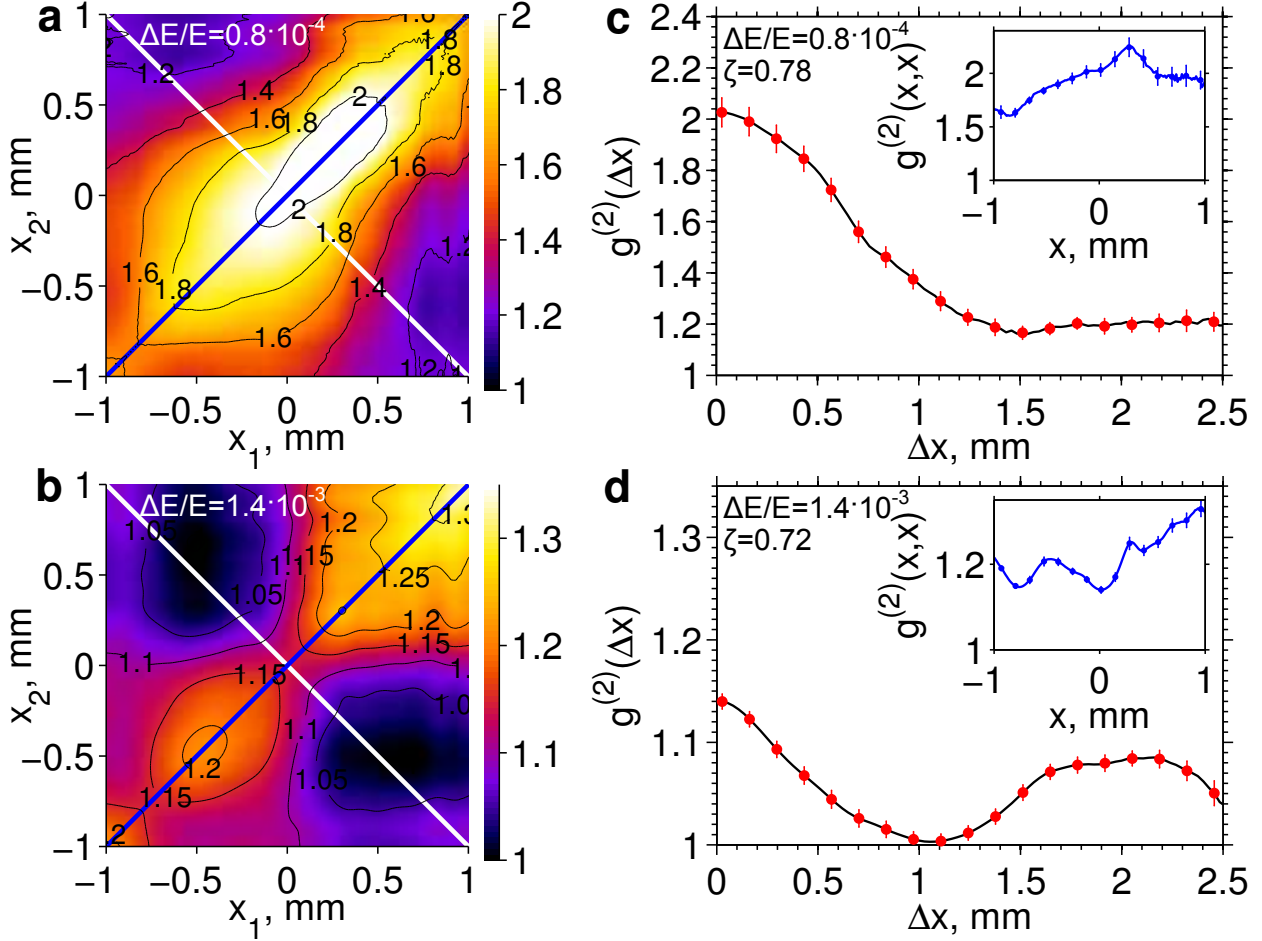}
\caption{(Color online) Intensity correlation analysis.
	(a,b) Intensity correlation function $g^{(2)}(x_1,x_2)$ for a bandwidth of $0.8\cdot10^{-4}$ (a) and $1.4\cdot10^{-3}$ (b).
	(c,d) Intensity correlation function $g^{(2)}(\Delta x)$ taken along the white line in (a,b) around the center of the beam.
	Insets in (c,d) show intensity fluctuations $g^{(2)}(x,x)=\avg{I^2(x)}/\avg{I(x)}^2$ taken along the blue line in (a,b).
	The error bars in (c,d) are obtained by statistical analysis of 20 individual sub-ensembles ($10^3$ shots each) from the whole set of $2\cdot10^4$ shots.
	The number of points displayed in this figure with error bars is reduced for better visibility.}
\label{fig:2}
\end{figure}
%
%
\begin{figure}[thb]
\centering
\includegraphics[width=1\linewidth]{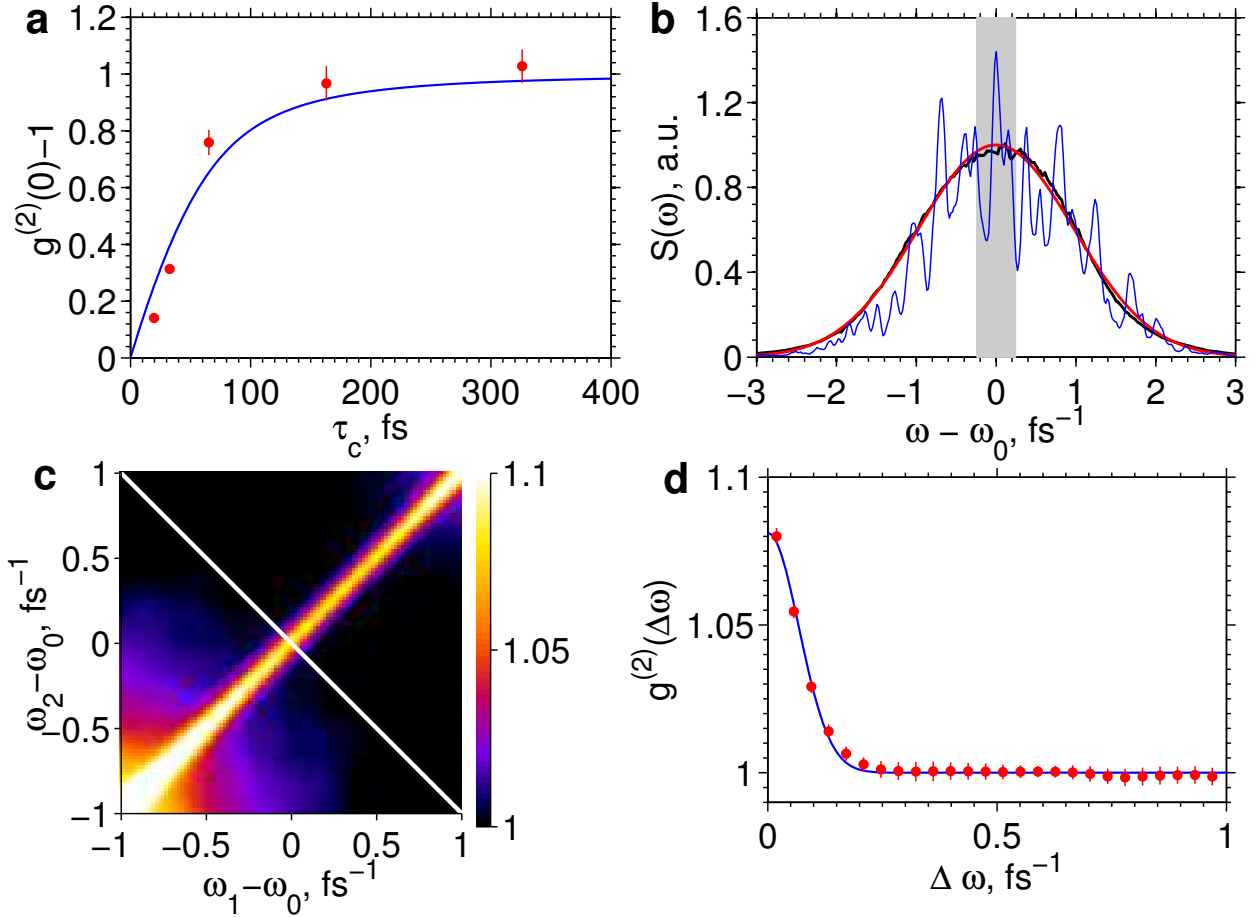}
\caption{(Color online) Statistical properties of FEL light as a function of the bandwidth.
	(a) Contrast (points) $\zeta_2 (D_\omega) = g^{(2)}(0)-1$ as a function the coherence time $\tau_c$.
	The error bars are same as in Fig. 2 (c,d).
	Theoretical fit (solid line) correspond to a pulse duration of $T=47\pm21$ fs.
	(b) Measured single pulse (blue line) and average (black line) spectra.
	Gaussian fit (red line) gives an FEL bandwidth of $\Delta E/E=6.7\cdot10^{-3}$ FWHM.
	The width of the largest exit slit used in the experiment and corresponding to the bandwidth $\Delta E/E=1.4\cdot10^{-3}$ is shown by the gray region.
	(c) Intensity correlation function $g^{(2)} (\omega_1,\omega_2)$ in the spectral domain.
	(d) Intensity correlation function $g^{(2)}(\Delta\omega)$ (red circles) taken along the white line in (c).
Gaussian fit (blue line) gives an average pulse duration of 27 fs.}
\label{fig:3}
\end{figure}
%
%
\begin{figure}[thb]
\centering
\includegraphics[width=1\linewidth]{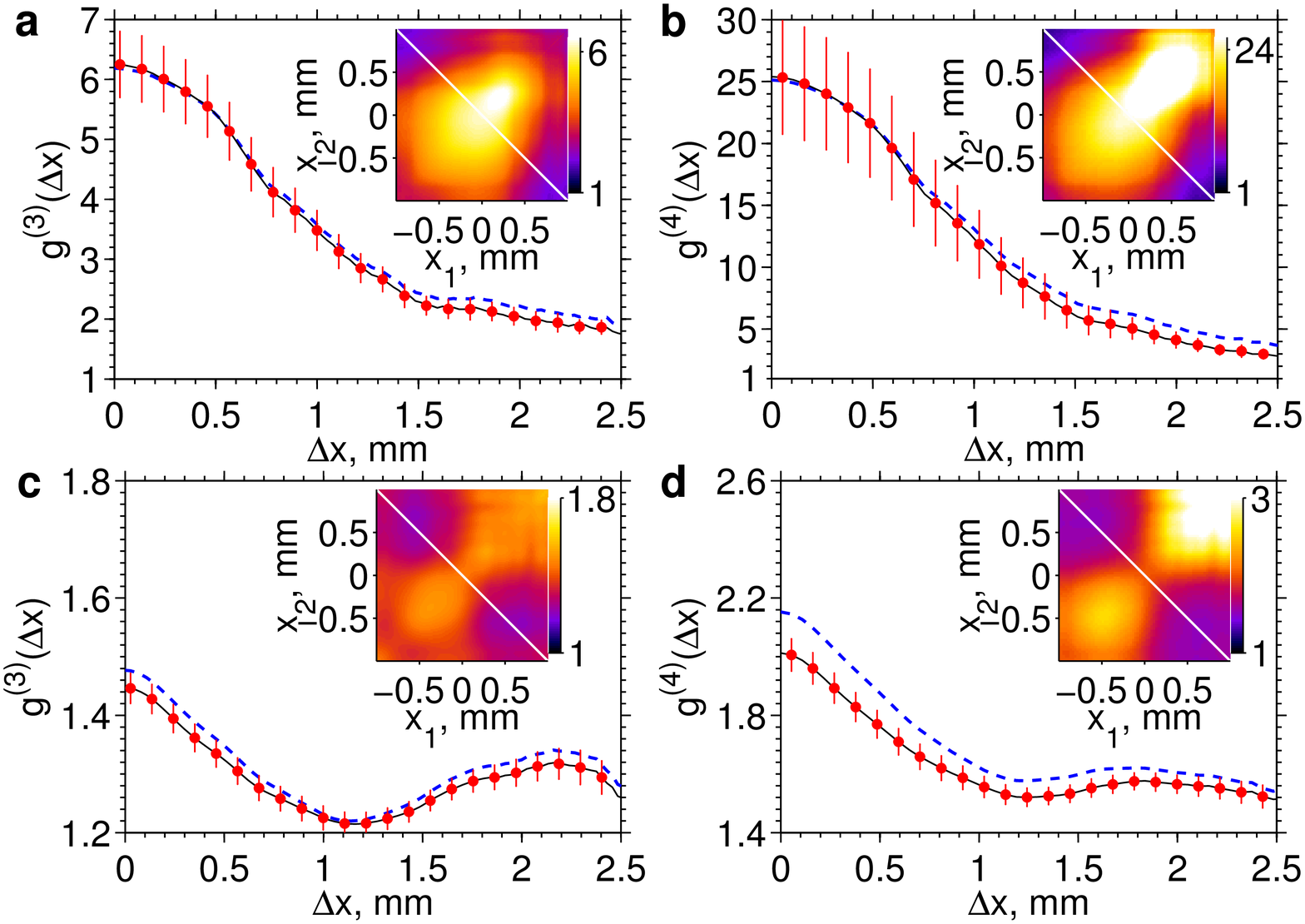}
\caption{(Color online) Higher order correlation functions for a bandwidth of $0.8\cdot10^{-4}$ (a, b) and $1.4\cdot10^{-3}$ (c, d).
	(a,c) Third-order correlation function $g^{(3)}(x_1, x_2)$ (shown in the insets) and $g^{(3)}(\Delta x)$ (black solid line) taken along the white line in the insets.
	(b,d) Fourth-order correlation function $g^{(4)}(x_1, x_2)$ (shown in the insets) and $g^{(4)}(\Delta x)$ (black solid line) taken along the white line in the insets.
	The same functions obtained under assumption of the Gaussian statistics are shown as blue dashed lines.
	The number of points displayed in this figure with error bars is reduced for better visibility.
	Error bars are evaluated in the same way as in Fig.~\ref{fig:2}.}
\label{fig:4}
\end{figure}

The experiment was carried out at FLASH that was operated with six undulator modules and a total undulator length of 30 m.
The electron bunch charge was 600 pC, and the electron energy 1.08 GeV resulting in a photon wavelength of $\lambda$=5.5 nm.
The average photon pulse energy was about 110 $\mu$J, which corresponds to about $3\cdot10^{12}$ photons per pulse at this photon energy.
The measurements were performed at the PG2 beam line \cite{35,36} (see Fig.~\ref{fig:1}).
The optical system focuses the beam at a distance of 71.5 m downstream from the undulator exit.
A monochromator comprised of a plane grating (PG), collimating (M1) and focusing (M2) mirrors, and an exit slit with variable slit width was utilized to modify the bandwidth.
The plane grating has a line density of 200 lines/mm and was tuned to its third order resulting in a dispersion in the exit slit plane of 0.64 eV/mm.
The resolution of the monocromator was ΔE=7 meV, which corresponds to an energy bandwidth of $\Delta E/E=3.1\cdot10^{-5}$.
Exit slits of 30 $\mu$m, 60 $\mu$m, 150 $\mu$m, 300 $\mu$m and 500 $\mu$m in size were used to select the spectral width.
These values correspond to energy bandwidths $\Delta E/E$ of $0.8\cdot10^{-4}$, $1.7\cdot10^{-4}$, $4.0\cdot10^{-4}$, $0.8\cdot10^{-3}$, and $1.4\cdot10^{-3}$.

In contrast to the original HBT experiment \cite{1,2} with two separated detectors, we utilize a pixel detector for correlation measurements.
It allows to determine second- and higher-order correlation functions at all separation points simultaneously.
An in-vacuum CCD (Andor Ikon, 2048 $\times$ 2048 pixels, each 13.5 $\mu$m $\times$ 13.5 $\mu$m in size) was positioned at a distance of 3.3 m behind the focus of the beamline.
The detector was operated at a repetition rate of 10 Hz with a region of interest 2047(H) $\times$ 460(V) pixels and binning by five pixels in the vertical direction.
A silicon nitride film 14 mm $\times$ 14 mm in size and 1 $\mu$m thick was positioned approximately 30 cm upstream from the camera to attenuate the beam.
The transmission of the film at this energy was $4\cdot10^{-6}$.
About $2\cdot10^4$ intensity profiles were recorded for each monochromator setting.
Dark images were subtracted from the measured intensity distribution, and occasionally occurring negative values were set to zero.
Typical single pulse and averaged intensities are shown in Fig.~\ref{fig:1} (b-d).
Shot to shot fluctuations in these intensity profiles, as a consequence of the SASE process are clearly visible.
The intensity profiles were averaged along the vertical direction, which is the dispersive direction of the monochromator (see Fig.~\ref{fig:1} (e-g)).
The intensity correlation analysis was performed in the horizontal direction.

The normalized second-order correlation function $g^{(2)}(x_1,x_2)$ for a narrow bandwidth of $0.8\cdot10^{-4}$ is shown in Fig.~\ref{fig:2} (a,c).
Remarkably, it reaches the maximum value of two at small separations (Fig.~\ref{fig:2} (c)), which indicates that the contrast $\zeta_2(D_\omega)$ at that monochromator setting is close to one.
This contrast is significantly higher than at synchrotron sources \cite{25,26,27}, where it did not exceed 0.3.
A Gaussian fit, $\exp⁡\left((-\Delta x^2)/(2l_c^2)\right)$, to the second-order correlation function  $g^{(2)} (\Delta x)$ as a function of the separation $\Delta x=x_2-x_1$ around the center of the beam (see Fig.~\ref{fig:2} (c)) provided a transverse coherence length of $l_c=0.93\pm0.04$ mm.
This value is substantially larger than the measured beam size in horizontal direction (0.45 mm (FWHM)), indicating a high coherence of the beam.
We quantified it by evaluating the degree of spatial coherence $\zeta_S$ defined as \cite{37,38} $\zeta_S=\int|W(x_1,x_2 )|^2 dx_1 dx_2 /(\int S(x)dx)^2$ and obtained $\zeta_S =0.78\pm0.01$, which is in a good agreement with the Young's double pinhole measurements at FLASH \cite{19}.
Importantly, we can estimate the degeneracy parameter \cite{14}, which is the number of photons in a single mode.
Our estimates, obtained by integrating the total flux on the detector at the narrow bandwidth, yield a value of $10^9$ comparable to Ref.~\cite{19}, which is significantly higher than at any synchrotron sources.

The second-order correlation function $g^{(2)}(x_1,x_2)$ for a larger bandwidth of $1.4\cdot10^{-3}$ is shown in Fig.~\ref{fig:2}(b,d).
This larger bandwidth is equivalent to a shorter coherence time, and the contrast of  $g^{(2)}(x_1,x_2)$ is reduced as expected from Eq.~\eqref{eq:2}.
Unexpectedly, at these conditions we observed an oscillatory behavior of the correlation function $g^{(2)} (\Delta x)$ (see Fig.~\ref{fig:2}(d)).
This may originate from the contribution of two independent sources in the lasing conditions of FLASH as discussed in Ref.~\cite{28}.
We want to remark that such fine features of FEL lasing would be difficult to observe by other means.
For example, our averaged spectral measurements (see Fig.~\ref{fig:3}(b)) do not show any notable features.

Results of our experiments also indicate that the statistical properties of the FEL beam are not spatially uniform.
This is well seen by the inspection of the beam fluctuations $g^{(2)}(x,x) =\avg{I^2(x)}/\avg{I(x)}^2$ along the beam profile (see insets in Fig.~\ref{fig:2}(c,d)).
They vary significantly both for small and large bandwidths.

We also analyzed the contrast in the center of the beam $\zeta_2 (D_\omega) = g^{(2)}(0)-1$ as a function of the coherence time $\tau_c$ (see Fig.~\ref{fig:3}(a)).
It has a linear dependence as a function of $\tau_c$ at a large bandwidth and reaches saturation at a small bandwidth, as described by Eq.~\eqref{eq:2}.
A theoretical fit to the contrast values yields an average pulse duration of $47\pm21$ fs (FWHM) (see the Appendix B for details)

To get an independent estimate of the pulse duration, we measured the second-order correlation function $g^{(2)}(\omega_1,\omega_2)$ in the spectral domain as a function of two frequencies (see Fig.~\ref{fig:3}(c)) \cite{29,30}.
About $1.5\cdot10^4$ single shot spectra were recorded after the intensity correlation measurements with a detector at the position of the exit slit of the monochromator.
This detector is comprised of a scintillating screen (YAG:Ce 0.2\%) and an intensified CCD (Andor iStar, DH740), equipped with a lens.
The effective pixel size of the detector in the exit slit plane was 19.4 $\mu$m with the point spread function estimated to be about two pixels (FWHM).
The detector was operated at a repetition rate of 10 Hz.
Line profiles, obtained from the central part of the beam in the horizontal direction corresponding to about 10 \% of the beam FWHM were analyzed.
These line profiles were background corrected by subtracting a constant offset, which was obtained from the averaged spectrum.
The occasionally occurring negative values in these line profiles were set to zero.
A Gaussian fit $\exp\left(-\Delta\omega^2\cdot T^2/2\right)$ was used to determine the pulse duration $T$ from the measurements of the second-order correlation function $g^{(2)}(\Delta\omega)$ (see Fig.~\ref{fig:3}(d)).
According to this analysis we obtained an average pulse duration of 27 fs (FWHM), which lies in the uncertainty range of the previous measurements.
In general, our estimates of the pulse duration can be shorter than the intrinsic pulse duration in time domain due to a possible frequency chirp of FLASH pulses \cite{30,31}.

To get an insight into the photon statistics of FEL pulses we studied intensity correlation functions of order higher than two.
Higher-order correlation functions  $g^{(3)}(x_1,x_2,x_3)$ (with $x_3=0$) and $g^{(4)}(x_1,x_2,x_3,x_4)$ (with $x_3=x_1/2$ and $x_4=x_2/2$) are presented in the insets of Fig.~\ref{fig:4} for two different bandwidths as functions of $x_1$ and $x_2$.
In Fig.~\ref{fig:4} the same quantities are presented as a function of $\Delta x$ around the center of the beam with $x_1=-\Delta x/2$, $x_2=\Delta x/2$.
The higher-order correlation functions were also calculated under the assumption of Gaussian statistics (see the Appendix A for details) from the measured second-order correlation function $g^{(2)}(x_1,x_2)$ (see Fig.~\ref{fig:4}).
The excellent agreement between these two curves confirms the chaotic nature of the FLASH source at these operation conditions.
In particular at a small bandwidth we observed a factorial behavior of higher order correlation functions at zero point separation $g^{(n)} (\br,\br,\ldots,\br)\approx n!$ (see Fig.~\ref{fig:4}(a,b)), which is typical for Gaussian statistics.
At a wider bandwidth the maximum value is substantially reduced compared to a narrow bandwidth case (see Fig.~\ref{fig:4}(c,d)).

In summary, we have presented intensity correlation measurements at FLASH which provide a simple, robust and versatile tool for monitoring basic beam properties of FELs including the degree of spatial coherence, average pulse duration, and details of the photon statistics.
We obtained an averaged pulse duration of 50 fs, degree of transverse coherence of about 80\%, and degeneracy parameter of the order of $10^9$.
Such values are similar to laser sources and were never observed at these wavelengths at conventional synchrotron radiation sources.
However, our measurements of the higher-order correlation functions indicate that present FEL sources based on the SASE principle are essentially chaotic sources obeying Gaussian statistics.

An interesting further application of the methods developed in this work would be the study of seeded FEL sources \cite{32,33}.
An intriguing question is whether seeded FELs are fully coherent sources in all orders according to Glauber \cite{3} and in this way are equivalent to conventional single mode lasers, or if they obey Gaussian statistics like SASE FELs.
We could also foresee that intensity correlation analysis might be applied to study the dynamics of ultrafast processes at FELs.
For example, the intensity correlation analysis of the Coulomb explosion in single molecule imaging experiments \cite{34} could provide detailed information about the disintegration of these molecules on a femtosecond time scale.

\begin{acknowledgments}
We acknowledge fruitful discussions with E. Saldin, E. Schneidmiller and M. Altarelli, the development of cross correlation analysis in frequency domain by S. Serkez, and careful reading of the manuscript by R. R\"ohlsberger.
Part of this work was supported by BMBF Grant No. 5K10CHG ''Coherent Diffraction Imaging and Scattering of Ultrashort Coherent Pulses with Matter'' in the framework of the German-Russian collaboration ''Development and Use of Accelerator-Based Photon Sources'' and the Virtual Institute VH-VI-403 of the Helmholtz Association.
The authors thank the FLASH machine and experiments team for their great support.
We also are greatly indebted to the scientific and technical team at FLASH, in particular the machine operators and run coordinators, being the foundation of the successful operation and delivery of the SASE-FEL beam.
\end{acknowledgments}

\appendix

\section{Higher order intensity correlation functions}

The normalized intensity correlation function is defined as
\begin{equation}
  g^{(2)}(\mathbf r_1,\mathbf r_2) =
  \frac{ \langle I( \mathbf r_1 ) I ( \mathbf r_2 ) \rangle }
  {\langle I ( \mathbf r_1 )\rangle \langle I (\mathbf r_2) \rangle},
  \label{eq:correlation2}
\end{equation}
where $\mathbf r_1$, $\mathbf r_2$ are positions in space.
In our implementation of the Hanbury Brown and Twiss experiment the intensity is given by
\begin{equation}
  I(\mathbf r)=\int\limits_{-\infty}^{\infty} |T(\mathbf r,\omega)|^2 |E(\mathbf r,\omega)|^2\mbox d\omega,
  \label{eq:intensity}
\end{equation}
where $E(\mathbf r,\omega)$ is the radiation field in the space-frequency domain and $T(\mathbf r,\omega)$ is the transmission function of the monochromator.
The average $\langle \cdots\rangle$ in Eq.~\eqref{eq:correlation2} is performed over an ensemble of different FEL pulses.
In our experiment the intensity correlation function $g^{(2)}(\mathbf r_1,\mathbf r_2)$ represents a coincidence measurement between different pixels of the detector.

Below we consider only the horizontal direction $x$ and a monochromator with a transmission function $T(\omega)$ that does not depend on the position.
Substituting Eq.~\eqref{eq:intensity} into Eq.~\eqref{eq:correlation2} and interchanging the order of average and integration we find
\begin{equation}
  g^{(2)}(x_1,x_2) =
  \frac{ \int\limits_{-\infty}^{\infty} \int\limits_{-\infty}^{\infty} |T(\omega_1)|^2|T(\omega_2)|^2\langle E^*( x_1 , \omega_1 ) E^*( x_2, \omega_2 )E( x_2 , \omega_2 ) E( x_1, \omega_1 ) \rangle \mbox d\omega_1 \mbox d\omega_2}
  { \int\limits_{-\infty}^{\infty} \int\limits_{-\infty}^{\infty}|T(\omega_1)|^2|T(\omega_2)|^2 \langle |E( x_1 , \omega_1 )|^2\rangle\cdot \langle |E( x_2, \omega_2 )|^2\rangle\mbox d\omega_1 \mbox d\omega_2 }
  \label{eq:correlation2_}
\end{equation}
For fields obeying Gaussian statistics, higher-order correlations of the field can be expressed through the first-order correlation functions using the Gaussian moment theorem \cite{14}
\begin{equation}
  \begin{split}
    \langle E^*(x_1,\omega_1)&\cdots E^*(x_n,\omega_n)E(x_n,\omega_n)\cdots E(x_1,\omega_1)\rangle\\
    &= \sum_\pi \langle  E^*(x_1,\omega_1) E^*(x_{\pi(1)},\omega_{\pi(1)})\rangle
    \cdots\langle E^*(x_n,\omega_n) E^*(x_{\pi(n)},\omega_{\pi(n)})\rangle.
    \label{eq:GaussianMomentTheorem}
  \end{split}
\end{equation}
Here $\pi$ denotes a summation over all $n!$ possible permutations of the subscripts.
Applying this theorem to Eq.~\eqref{eq:correlation2_} with $n=2$ yields
\begin{equation}
  \begin{split}
  g^{(2)}(x_1,x_2)
=&1+\frac{\int\limits_{-\infty}^{\infty}\int\limits_{-\infty}^{\infty}|T(\omega_1)|^2|T(\omega_2)|^2 |W^{(1)}(x_1,x_2,\omega_1,\omega_2)|^2\mbox d\omega_1 \mbox d\omega_2}
 {\int\limits_{-\infty}^{\infty}\int\limits_{-\infty}^{\infty}|T(\omega_1)|^2|T(\omega_2)|^2 S(x_1,\omega_1)S(x_2,\omega_2)\mbox d\omega_1 \mbox d\omega_2},
  \label{eq:S5}
\end{split}
\end{equation}
where
\begin{equation}
  W^{(1)}(x_1,x_2,\omega_1,\omega_2)=
  \langle E^*(x_1,\omega_1) E(x_2,\omega_2)\rangle
    \label{eq:S6}
\end{equation}
is the first-order correlation function in space-frequency domain, or cross spectral density \cite{14} and
\begin{equation}
  S(x,\omega) = W^{(1)}(x,x,\omega,\omega)
    \label{eq:S7}
\end{equation}
is the spectral density.

With the assumption that the cross spectral density  and spectral density can be separated into its spatial and spectral part
\begin{equation}
  \begin{split}
   W^{(1)}(x_1,x_2,\omega_1,\omega_2)&=W(x_1, x_2)W(\omega_1, \omega_2), \\
   S(x,\omega)& =S(x) S(\omega)
 \end{split}
   \label{eq:separability}
\end{equation}
the intensity correlation function reduces to the form
\begin{equation}
  \begin{split}
  g^{(2)}(x_1,x_2)
    =&1 + \zeta_2(D_{\omega}) \cdot |\mu(x_1,x_2)|^2,
\end{split}
  \label{eq:correlation2_final}
\end{equation}
where
\begin{equation}
  \mu(x_1,x_2) = \frac{W(x_1,x_2)}{\sqrt{S(x_1)S(x_2) }}
    \label{eq:S10}
\end{equation}
is the normalized spectral degree of coherence.
In Eq.~\eqref{eq:correlation2_final} the contrast $\zeta_2(D_{\omega})$ of the second-order correlation function is defined as
\begin{equation}
  \zeta_2(D_{\omega}) = \frac{\int\limits_{-\infty}^{\infty}\int\limits_{-\infty}^{\infty} |T(\omega_1)|^2|T(\omega_2)|^2|W(\omega_1,\omega_2)|^2 \mbox d\omega_1 \mbox d\omega_2}
    {\Bigl(\int\limits_{-\infty}^{\infty}|T(\omega)|^2S(\omega)\mbox d\omega \Bigr)^2}
    \label{eq:zeta_2}
\end{equation}
and $D_{\omega}$ is the spectral width of the monochromator.
The functions $\mu(\omega_1,\omega_2)$, and $S(\omega)$ are defined accordingly from $W(\omega_1,\omega_2)$.

The third-order correlation function is given by
\begin{equation}
  g^{(3)}(x_1,x_2,x_3) =
  \frac{ \langle I( x_1 ) I ( x_2 ) I(x_3)\rangle }
  {\langle I ( x_1 )\rangle \langle I (x_2) \rangle\langle I(x_3)\rangle}.
  \label{eq:correlation3}
\end{equation}
Substituting Eq.~\eqref{eq:intensity} into Eq.~\eqref{eq:correlation3}, applying Gaussian moment theorem (\ref{eq:GaussianMomentTheorem}), and using approximation (\ref{eq:separability}) we find
\begin{equation}
  \begin{split}
    g^{(3)}(x_1,x_2,x_3) =& 1 + \zeta_2(D_{\omega}) \cdot \Bigl(|\mu(x_1,x_2)|^2 + |\mu(x_2,x_3)|^2 + |\mu(x_3,x_1)|^2\Bigr)\\
    + &2\zeta_3(D_{\omega}) \cdot \textrm{Re}\Bigl(\mu(x_1,x_2)\mu(x_2,x_3)\mu(x_3,x_1)\Bigr)
  \end{split}
  \label{eq:correlation3_final}
\end{equation}
where $\zeta_2(D_{\omega})$ has been defined in \eqref{eq:zeta_2} and the contrast $\zeta_3(D_{\omega})$  for the three-point correlation function is given by
\begin{equation}
\zeta_3(D_{\omega}) = \frac{\int\limits_{-\infty}^{\infty}\int\limits_{-\infty}^{\infty}\int\limits_{-\infty}^{\infty} |T(\omega_1)|^2|T(\omega_2)|^2|T(\omega_3)|^2 W(\omega_1,\omega_2) W(\omega_2,\omega_3) W(\omega_3,\omega_1) \mbox d\omega_1 \mbox d\omega_2 \mbox d\omega_3}
{\Bigl(\int\limits_{-\infty}^{\infty} |T(\omega)|^2S(\omega)\mbox d\omega\Bigr)^3}.
    \label{eq:zeta_3}
\end{equation}

In a similar way the fourth-order correlation function
\begin{equation}
  g^{(4)}(x_1,x_2,x_3,x_4) =
  \frac{ \langle I( x_1 ) I ( x_2 ) I(x_3) I(x_4)\rangle }
  {\langle I ( x_1 )\rangle \langle I (x_2) \rangle\langle I(x_3)\rangle\langle I(x_4)\rangle}
  \label{eq:correlation4}
\end{equation}
can be expressed through the first-order correlation functions in the frame of Gaussian statistics
\begin{equation}
  \begin{split}
	g^{(4)}&(x_1,x_2,x_3,x_4) = 1 + \zeta_2(D_{\omega}) \cdot \Bigl( |\mu(x_1,x_2)|^2 + |\mu(x_1,x_3)|^2 + |\mu(x_1,x_4)|^2 \Bigr)\\
    &+ \zeta_2(D_{\omega}) \cdot \Bigl( |\mu(x_2,x_3)|^2 + |\mu(x_2,x_4)|^2 + |\mu(x_3,x_4)|^2 \Bigr)  \\
		&+ \zeta_2(D_{\omega})^2 \cdot\Bigl( |\mu(x_1,x_2)|^2 |\mu(x_3,x_4)|^2 + |\mu(x_1,x_3)|^2 |\mu(x_2,x_4)|^2 + |\mu(x_1,x_4)|^2 |\mu(x_2,x_3)|^2 \Bigr)  \\
    &+ 2\zeta_3(D_{\omega}) \cdot\textrm{Re} \Bigl( \mu(x_2,x_3)\mu(x_3,x_4)\mu(x_4,x_2) + \mu(x_1,x_3)\mu(x_3,x_4)\mu(x_4,x_1) \Bigr)\\
		&+ 2\zeta_3(D_{\omega}) \cdot\textrm{Re} \Bigl( \mu(x_1,x_2)\mu(x_2,x_4)\mu(x_4,x_1) + \mu(x_1,x_2)\mu(x_2,x_3)\mu(x_3,x_1) \Bigr)  \\
    &+ 2\zeta_4(D_{\omega}) \cdot\textrm{Re} \Bigl( \mu(x_1,x_2)\mu(x_2,x_3)\mu(x_3,x_4)\mu(x_4,x_1)\Bigr)\\
    &+ 2\zeta_4(D_{\omega}) \cdot\textrm{Re} \Bigl( \mu(x_1,x_3)\mu(x_3,x_2)\mu(x_2,x_4)\mu(x_4,x_1)\Bigr)\\
    &+ 2\zeta_4(D_{\omega}) \cdot\textrm{Re} \Bigl( \mu(x_1,x_3)\mu(x_3,x_4)\mu(x_4,x_2)\mu(x_2,x_1) \Bigr),
  \end{split}
  \label{eq:correlation4_final}
\end{equation}
where $\zeta_2(D_{\omega})$ and $\zeta_3(D_{\omega})$ have the same meaning as in Eqs.~\eqref{eq:zeta_2} and \eqref{eq:zeta_3}.
The contrast $\zeta_4(D_{\omega})$ for the four-point correlation function is defined as
\begin{equation}
	\zeta_4(D_{\omega}) = \frac{\int f(\omega_1, \omega_2) f(\omega_2, \omega_3) f(\omega_3, \omega_4) f(\omega_4, \omega_1) d\omega_1 d\omega_2 d\omega_3 d\omega_4}
		{\Bigl( \int |T(\omega)|^2S(\omega) \mbox d\omega \Bigr)^4},
\end{equation}
where $f(\omega_i,\omega_j)=T^*(\omega_i)T(\omega_j) W(\omega_i,\omega_j)$.
The quantities $\zeta_2(D_{\omega})$, $\zeta_3(D_{\omega})$, and $\zeta_4(D_{\omega})$ are real numbers.
The phase of the cyclic product of SDC in Eqs.~\eqref{eq:correlation3_final} and \eqref{eq:correlation4_final} vanishes in most practical cases, and was thus neglected here.
This includes fully coherent and incoherent beams, and Gaussian Schell-model beams \cite{14}.

To investigate whether FEL radiation obeys Gaussian statistics, we determined the modulus of the spectral degree of coherence $|\mu(x_1,x_2)|$ and contrast $\zeta_2(D_{\omega})$ from the second-order correlation function measurements according to Eq.~\eqref{eq:correlation2_final}.
We then compared the third-order correlation function determined from the experiment (see Eq.~\eqref{eq:correlation3}) with the expression \eqref{eq:correlation3_final}.
The same was done for the fourth-order correlation function determined from the experiment using Eq.~\eqref{eq:correlation4} and compared with Eq.~\eqref{eq:correlation4_final}.
The parameters $\zeta_3(D_{\omega})$ and $\zeta_4(D_{\omega})$ in Eqs.~\eqref{eq:correlation3_final} and \eqref{eq:correlation4_final} were calculated as a function of $\zeta_2(D_{\omega})$ assuming a Gaussian form of the monochromator transmission function $T(\omega)=\exp\left( -\omega^2/[4 \cdot D_{\omega}^2]\right)$ and Gaussian Schell-model pulses \cite{LajunenJOSA2005}.
In the frame of this model we obtained
\begin{equation}
  \zeta_3(D_{\omega}) = \frac{4 \zeta_2(D_{\omega})^2}{ 3+\zeta_2(D_{\omega})^2}\qquad \textrm{and}\qquad \zeta_4(D_{\omega}) = \frac{2\zeta_2(D_{\omega})^3}{1+\zeta_2(D_{\omega})^2}.
\end{equation}

\section{Determination of the pulse duration from the intensity correlation measurements}

To determine the pulse duration we analyzed the contrast $\zeta_2(D_{\omega})$ obtained from the intensity correlation  measurements performed with different monochromator settings.
The transmission function of the monochromator was considered in the following form
\begin{equation}
  T(\omega)=\begin{cases}1&\textrm{for} \quad |\omega| \leq D_\omega/2; \\ 0&\textrm{for} \quad |\omega| > D_\omega/2 \end{cases}.
\end{equation}
The FEL pulses before the monochromator were taken in the form of Gaussian Schell-model pulses \cite{LajunenJOSA2005}.
In this model the spectral part of the cross spectral density $W(\omega_1,\omega_2)$ (see Eq.~\eqref{eq:separability}) is given by \cite{LajunenJOSA2005}
\begin{equation}
  W(\omega_1,\omega_2) =
  \exp\left[
  -\frac{ (\omega_1 - \omega_0)^2 + (\omega_2 - \omega_0)^ 2 } { 4 \Omega ^ 2 }
  -\frac{ ( \omega_2 - \omega_1 ) ^ 2 } { 2 \Omega_c ^ 2 }
  \right],
\end{equation}
where $\Omega$ is the spectral width of the pulse before the monochromator, $\Omega_c$ is the correlation width of the spectrum, and $\omega_0$ is the mean angular frequency.
These parameters can be expressed through the r.m.s. values of the pulse duration $T$ and the coherence time $T_c$ of the pulse before the monochromator \cite{LajunenJOSA2005}
\begin{equation}
  \Omega ^ 2 = \frac 1 { T_c ^ 2 } + \frac 1 { 4T ^ 2 } \quad \textrm{ and } \quad
    \Omega_c = \frac{T_c}{T}\Omega.
\end{equation}

In the conditions of our experiment the coherence time $T_c$ before the monochromator is much smaller than the pulse duration $T$.
Then the correlation width, $\Omega_c$,  in the spectral domain is well approximated by
\begin{equation}
  \Omega_c \approx \frac{1}{T}.
\end{equation}
Under the assumption that the transmitted bandwidth is much narrower than the bandwidth of the incoming FEL radiation $D_\omega\ll\Omega$, (see Fig.~3(b) of the main text) the contrast $\zeta_2(D_\omega)$ as a function of the transmitted bandwidth $D_\omega$ can be calculated by integrating Eq.~\eqref{eq:zeta_2} analytically
\begin{equation}
  \zeta_2(D_\omega) = \frac{\sqrt{\pi}}{D_\omega T} \textrm{erf}\left( D_\omega T \right)
  + \frac{1}{(D_\omega T)^2}\left\{\exp\left[-( D_\omega T ) ^ 2\right] - 1 \right\}.
\label{eq:fitting}
\end{equation}
According to Eq.~\eqref{eq:fitting} the contrast $\zeta_2(D_\omega)$ approaches the value of one for $D_\omega T\ll1$
and it has a linear behavior as a function of $1/D_\omega$ when condition $D_\omega T\gg1$ is satisfied. According to the definition of the coherence time $\tau_c=2\pi/D_\omega$ we obtain in this limit $\zeta_2(D_\omega)\sim\tau_c/T$.

The pulse duration was determined by fitting Eq.~\eqref{eq:fitting} to the contrast $\zeta_2(D_\omega)$ for different measurements, corresponding to different bandwidths $D_\omega$ of the monochromator settings (see Fig.~3(a) of the main text).

\section{Results of the measurements from all monochromator settings}

The results from all measured monochromator settings are summarized and presented in Table 1.

\begin{table}[h]
  \textbf{Table 1}\linebreak
  \phantom{a}\\
  \begin{tabular}{|c|c|c|c|c|c|}
  \hline
  $\Delta E/E$      &  Beam FWHM  & Coherence Length  & Degree of coherence & Contrast      & Number of modes\\
  \hline
$0.8\cdot 10^{-4}$  & 0.45 mm     & $0.93\pm0.04$ mm  & $0.78\pm0.01$       & $1.03\pm0.05$ & $1.5\pm0.1$\\
$1.7\cdot 10^{-4}$  & 0.45 mm     & $0.93\pm0.03$ mm  & $0.77\pm0.01$       & $0.97\pm0.06$ & $1.7\pm0.1$\\
$4.0\cdot 10^{-4}$  & 0.46 mm     & $1.13\pm0.06$ mm  & $0.89\pm0.01$       & $0.76\pm0.04$ & $1.4\pm0.1$\\
$0.8\cdot 10^{-3}$  & 0.46 mm     & $0.86\pm0.03$ mm  & $0.82\pm0.01$       & $0.31\pm0.01$ & $3.5\pm0.2$\\
$1.4\cdot 10^{-3}$  & 0.46 mm     & $0.53\pm0.04$ mm  & $0.72\pm0.01$       & $0.14\pm0.01$ & $7.8\pm0.4$\\
\hline
\end{tabular}
\end{table}
\bigskip

\begin{figure}[tp]
  \centering
  \includegraphics[height=0.15\textheight]{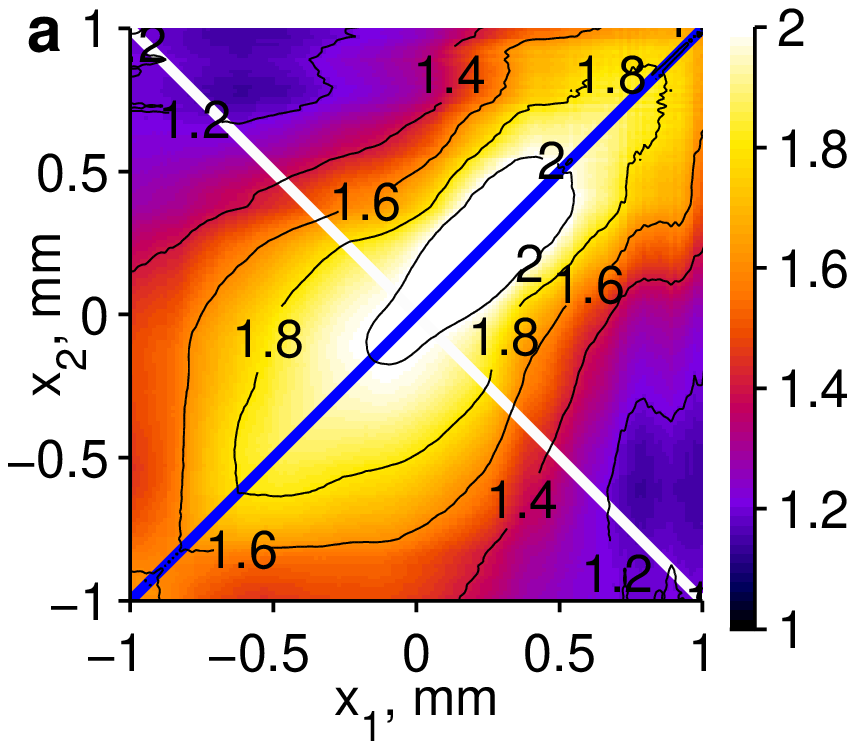}
  \includegraphics[height=0.15\textheight]{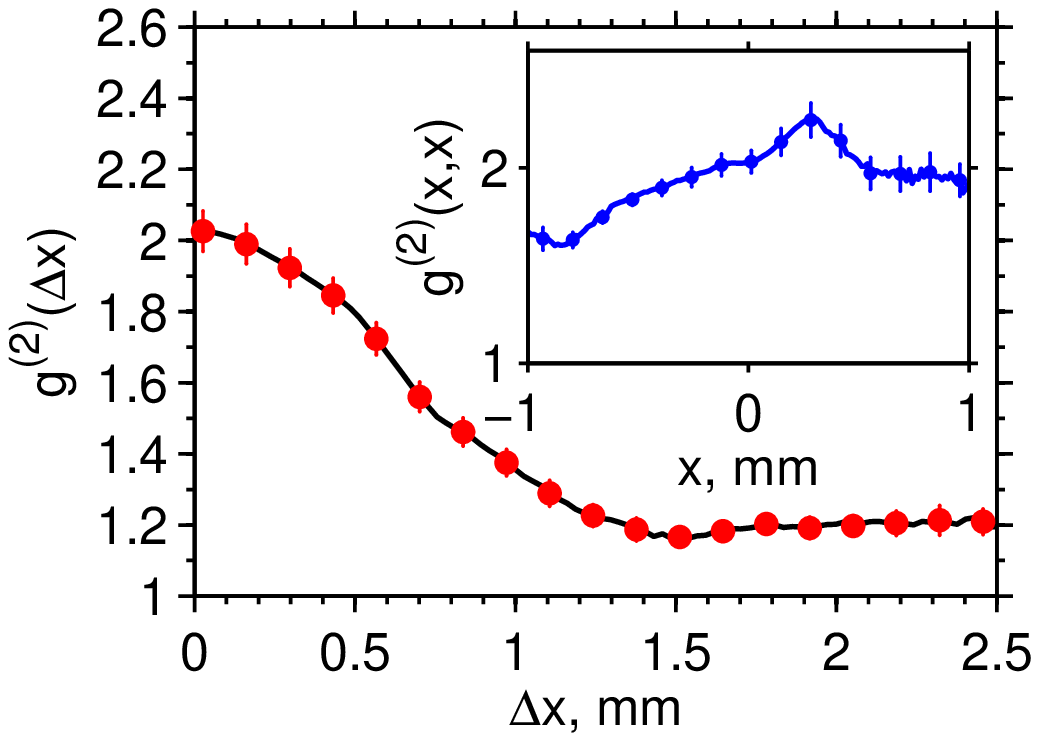}
  \includegraphics[height=0.15\textheight]{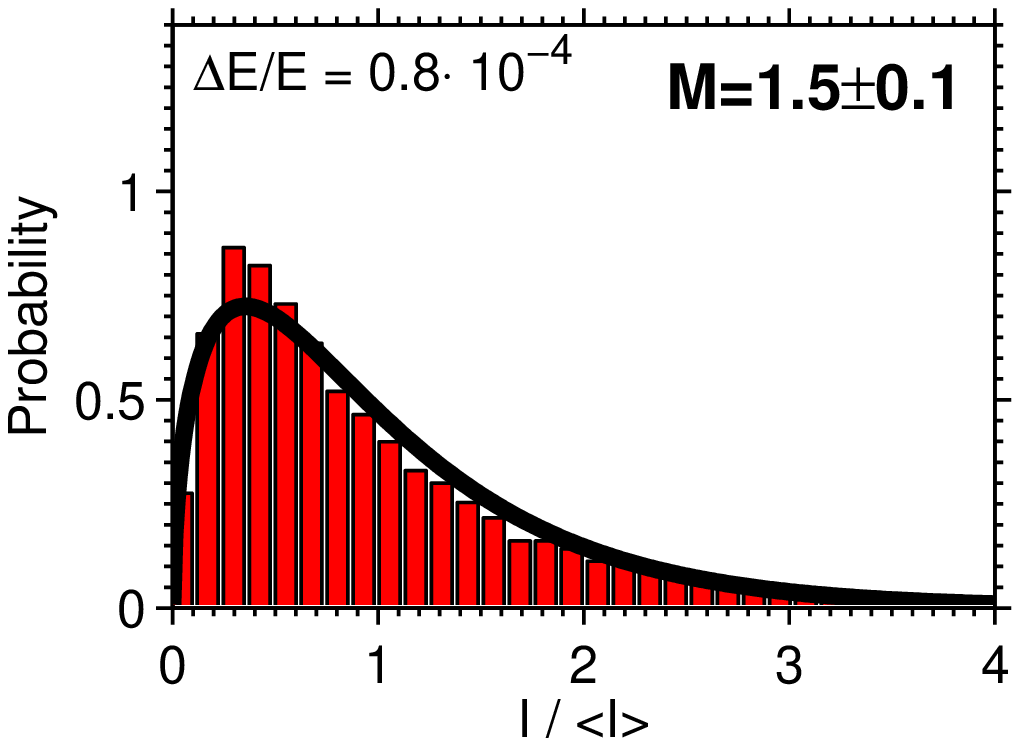}\\
  \includegraphics[height=0.15\textheight]{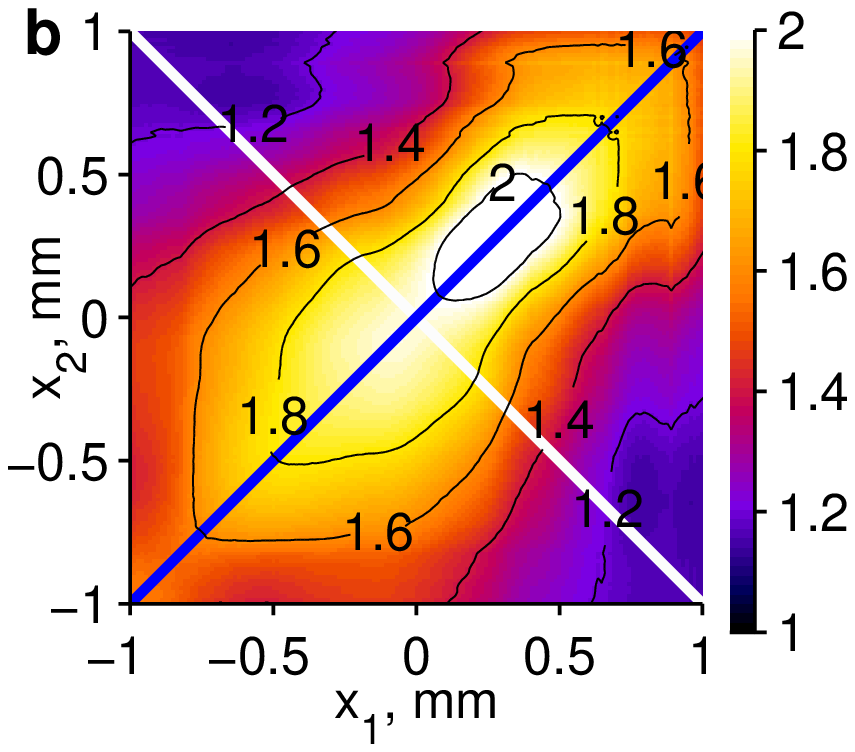}
  \includegraphics[height=0.15\textheight]{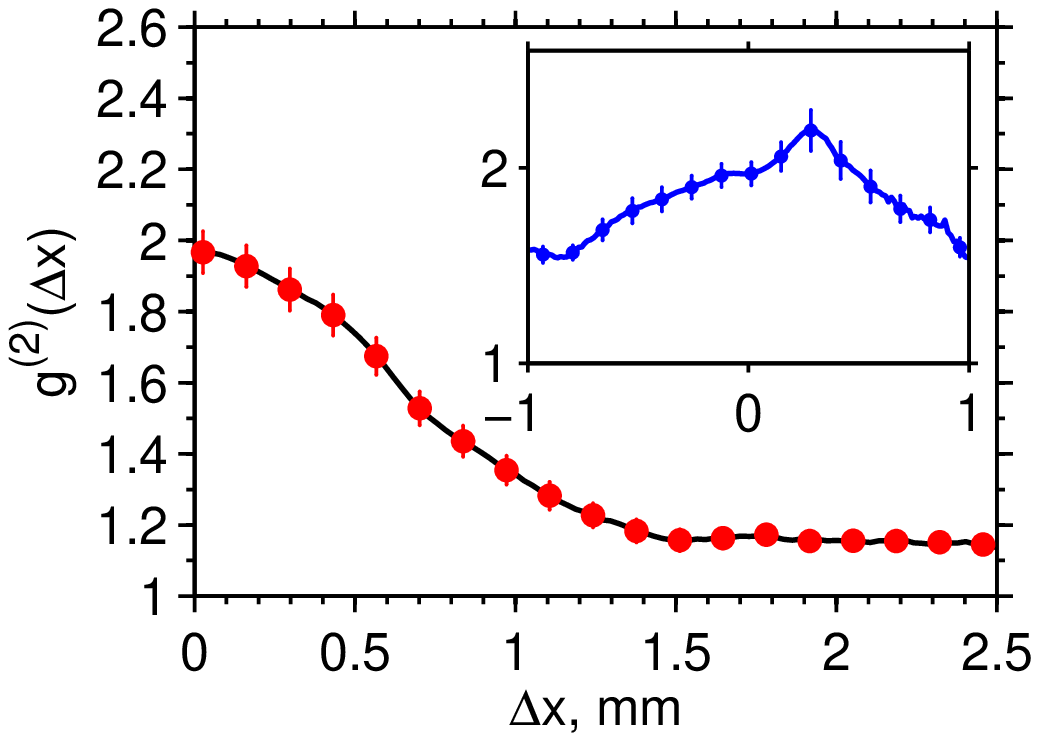}
  \includegraphics[height=0.15\textheight]{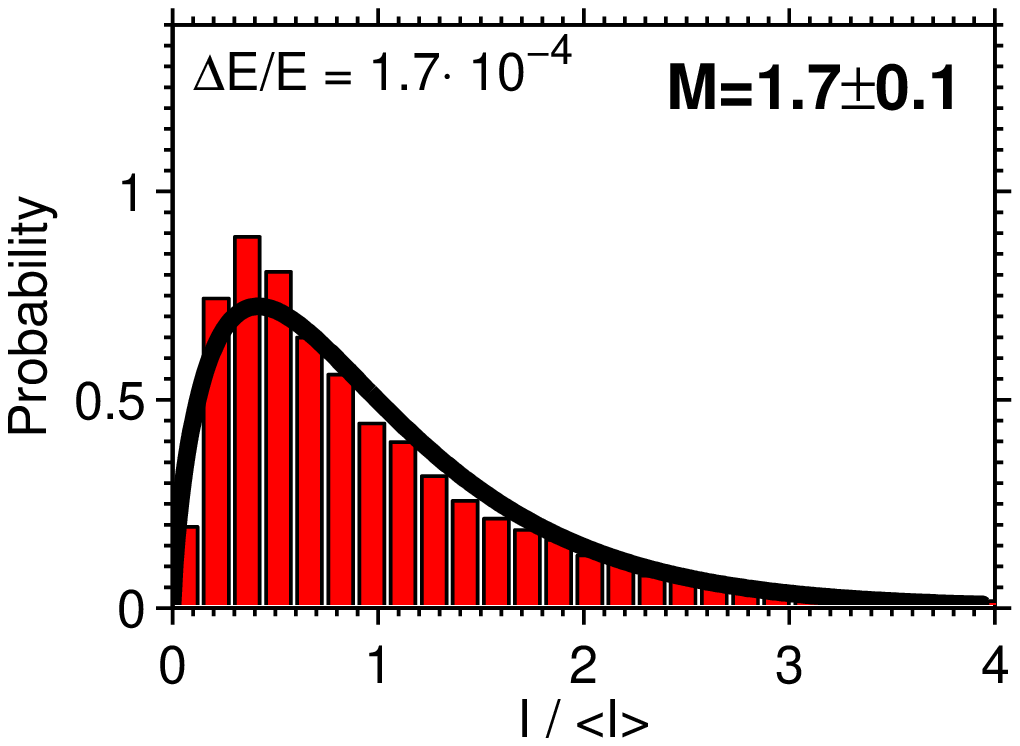}\\
  \includegraphics[height=0.15\textheight]{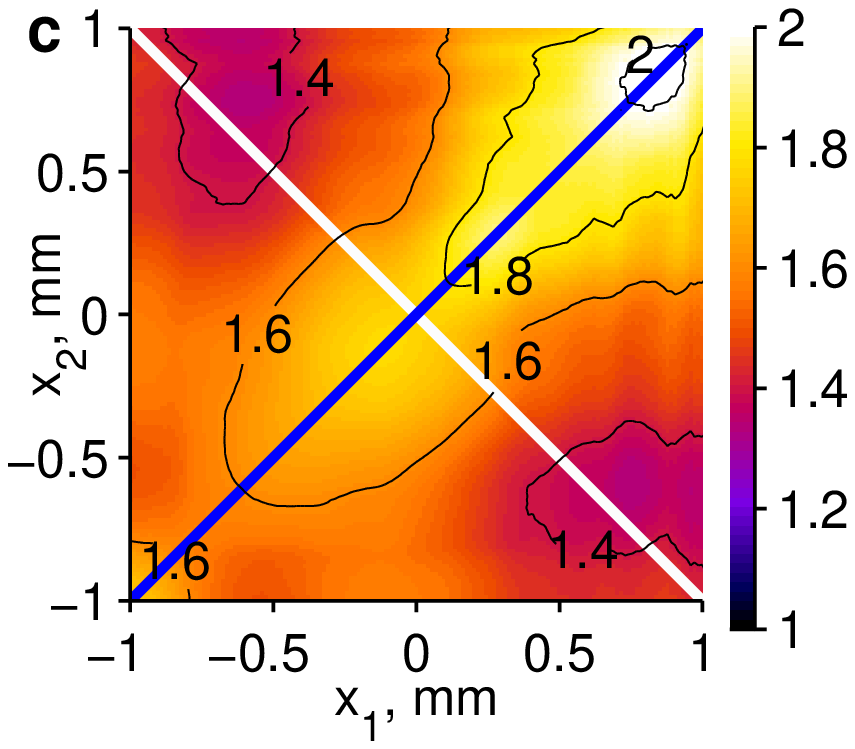}
  \includegraphics[height=0.15\textheight]{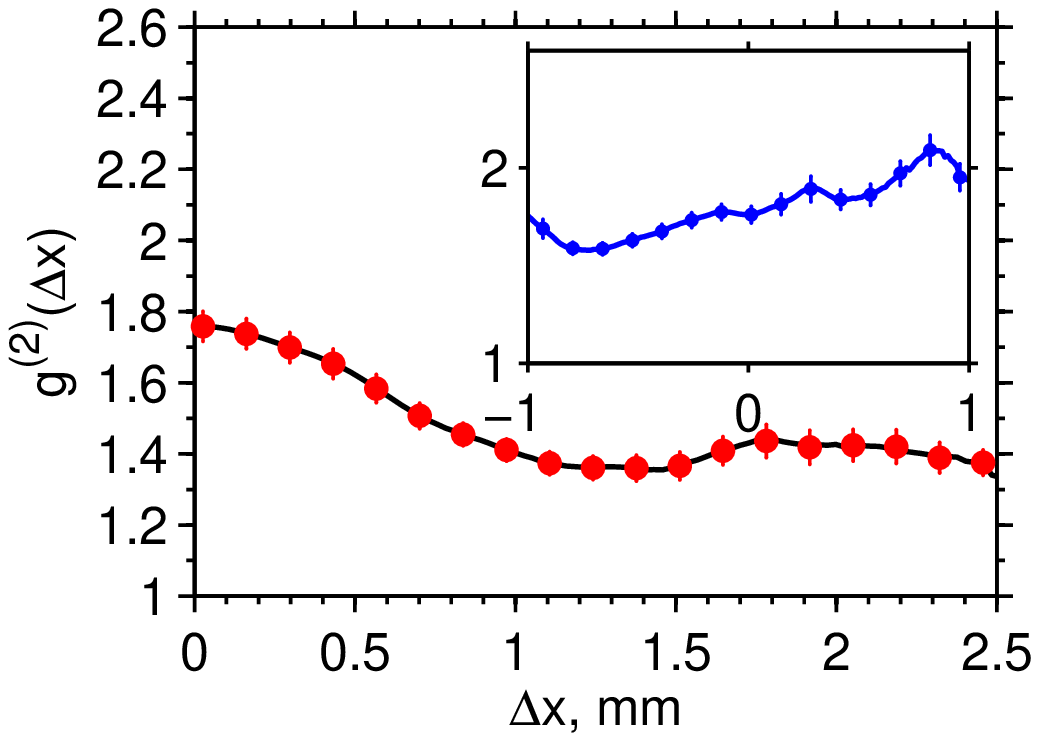}
  \includegraphics[height=0.15\textheight]{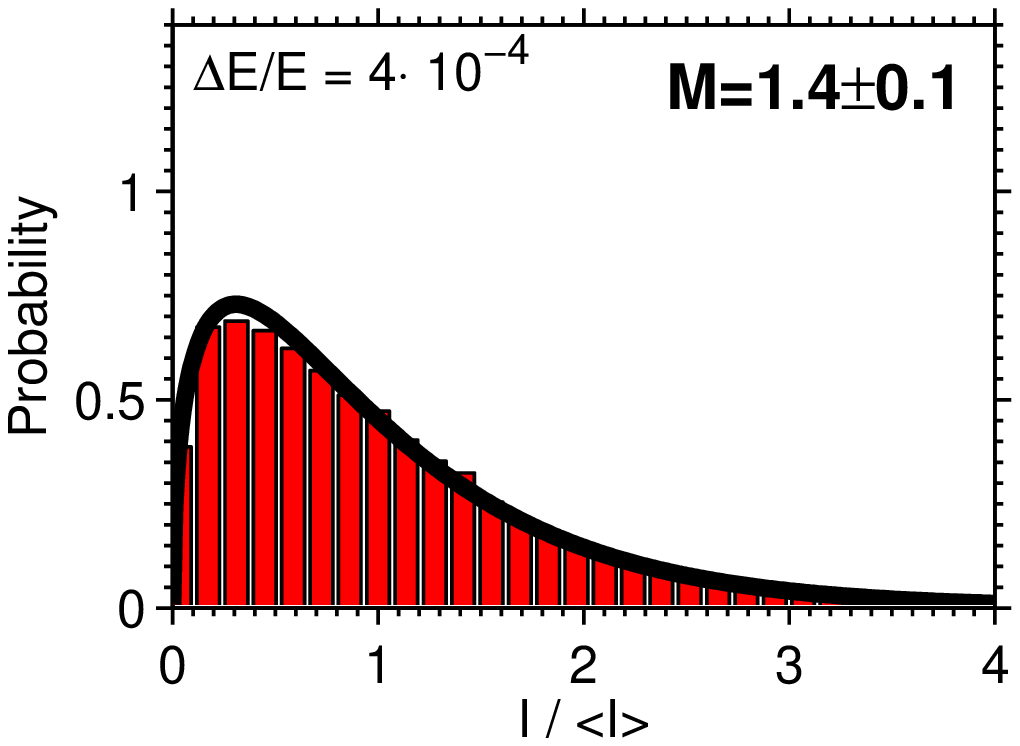}\\
  \includegraphics[height=0.15\textheight]{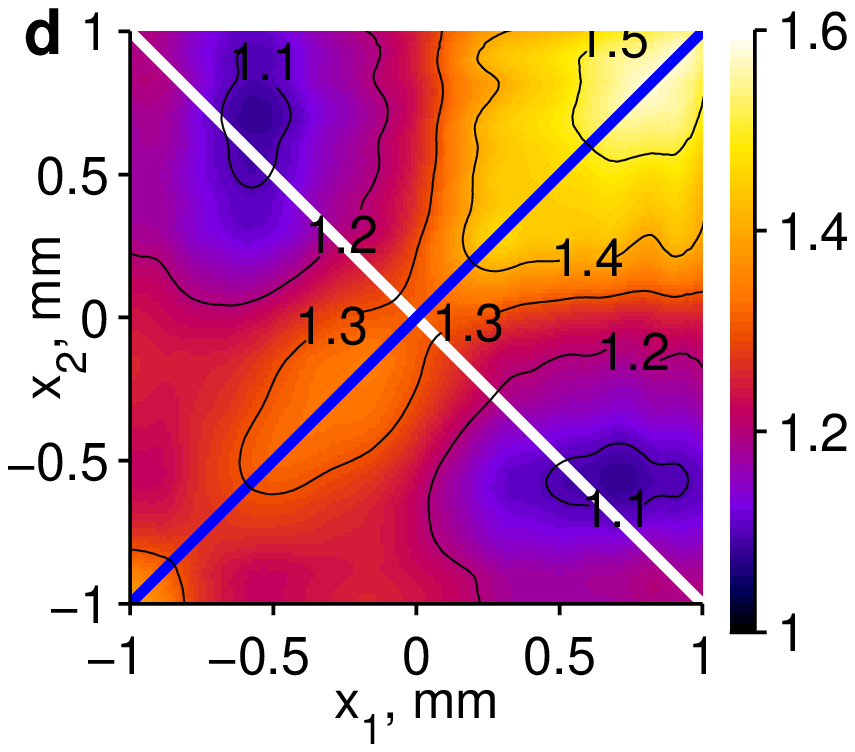}
  \includegraphics[height=0.15\textheight]{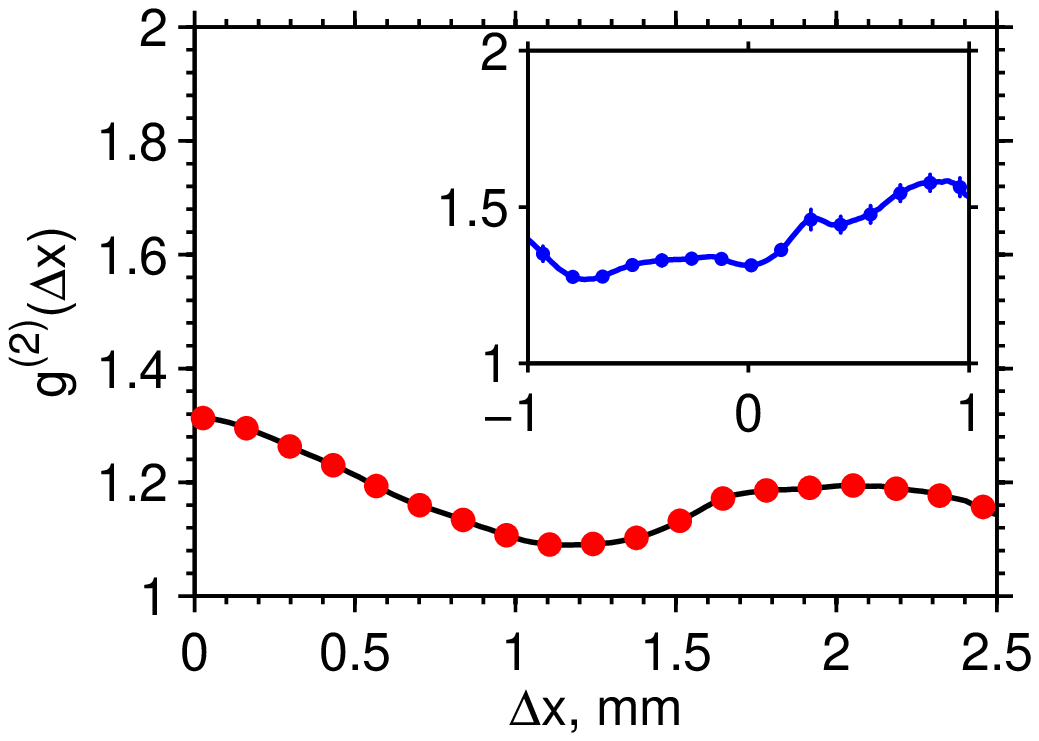}
  \includegraphics[height=0.15\textheight]{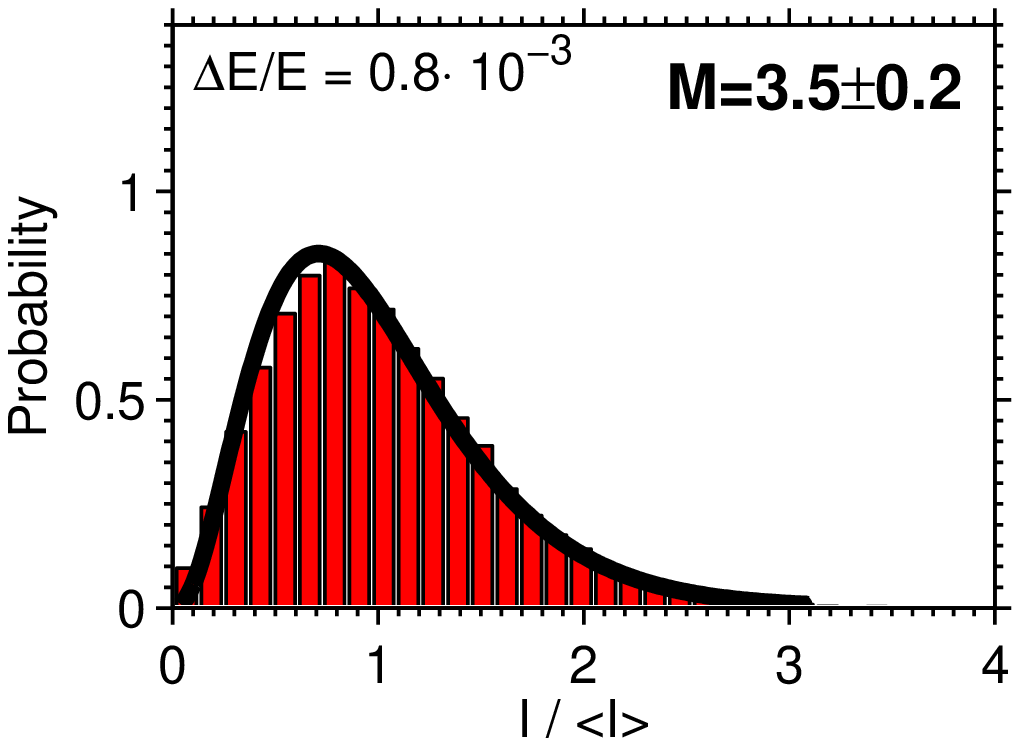}\\
  \includegraphics[height=0.15\textheight]{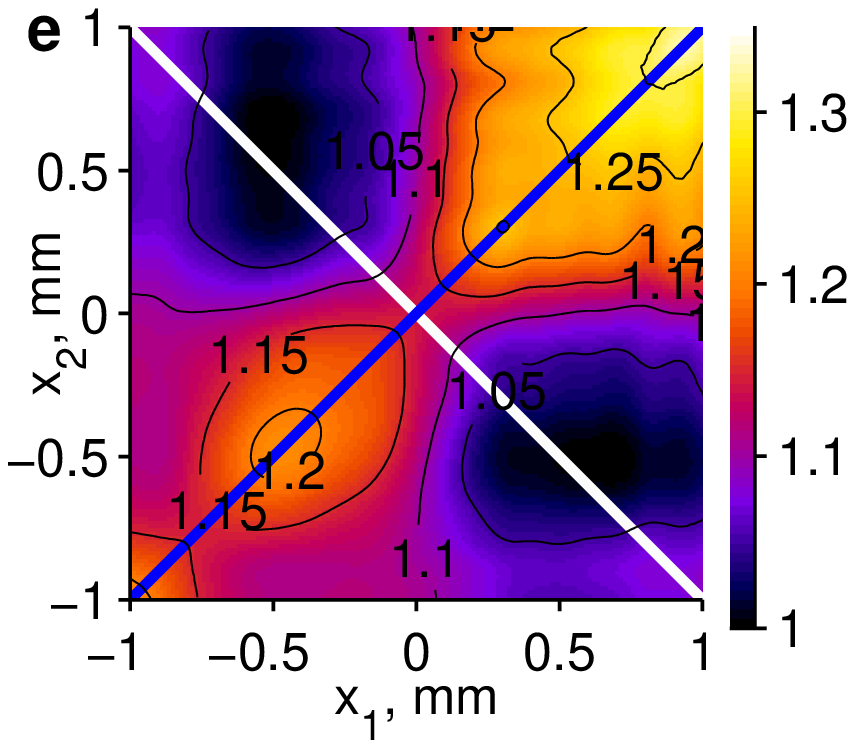}
  \includegraphics[height=0.15\textheight]{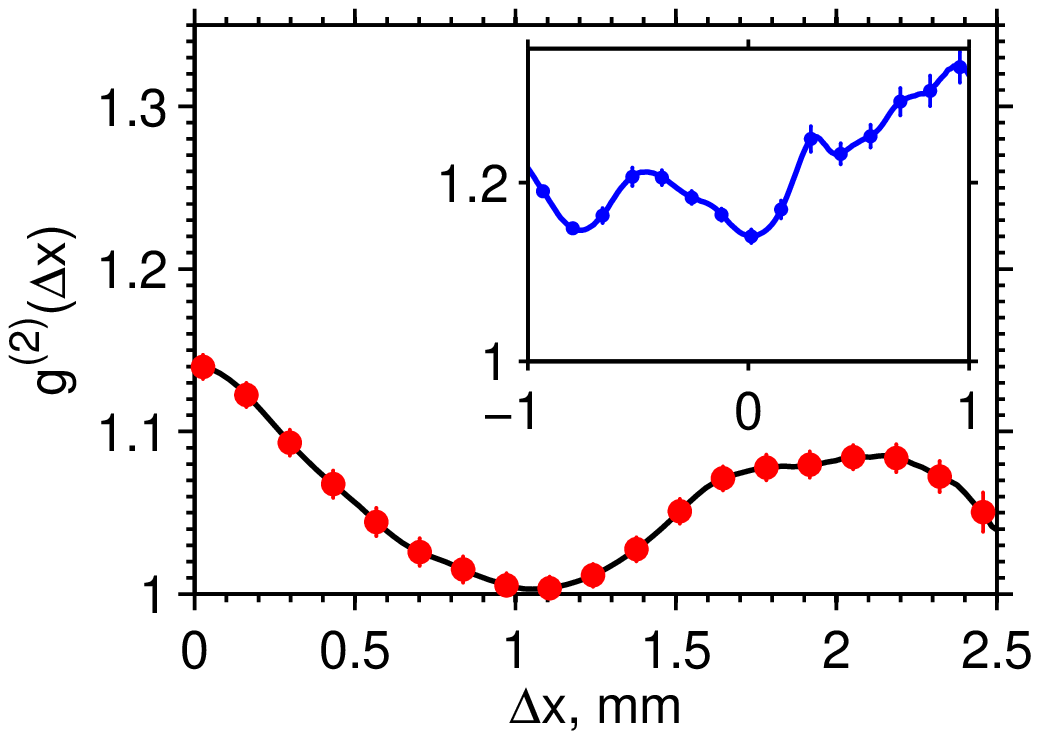}
  \includegraphics[height=0.15\textheight]{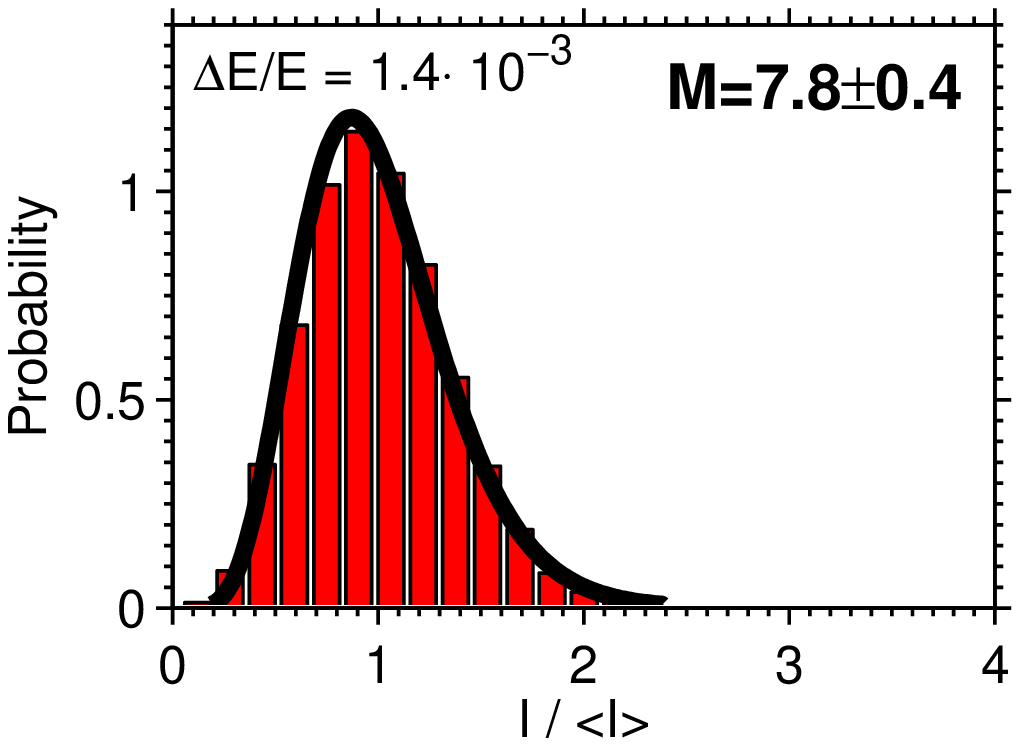}\\
  \caption{%
  Left and middle column: Same as in Fig.~\ref{fig:2} of the main text for all measured monochromator settings.
  The monochromator bandwidth $\Delta E/E$ was (a) $0.8\cdot10^{-4}$, (b) $1.7\cdot10^{-4}$, (c) $4\cdot10^{-4}$, (d) $8\cdot10^{-3}$, and (e) $1.4\cdot10^{-3}$.
  Right column: Histograms of the total intensity per pulse (red bars).
  Theoretical fits with the gamma function \cite{37} (black lines) and the number of total modes M (spatial and temporal) are also shown.
  }
  \label{fig:S1}
\end{figure}

\begin{figure}[tp]
  \centering
  \includegraphics[height=0.15\textheight]{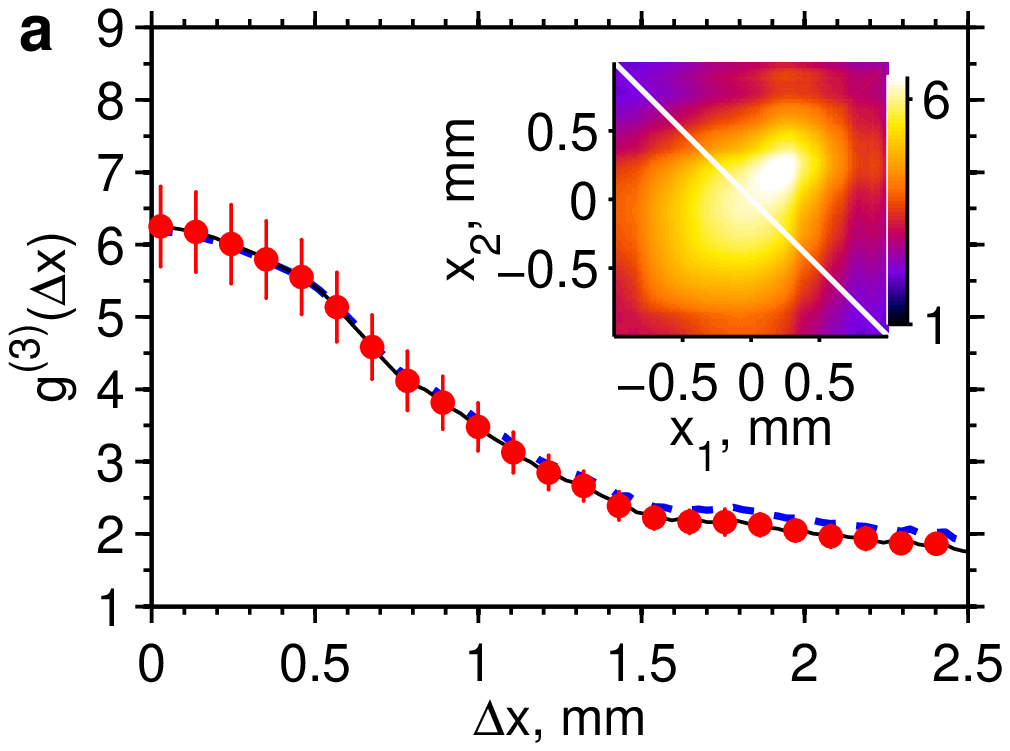}
  \includegraphics[height=0.15\textheight]{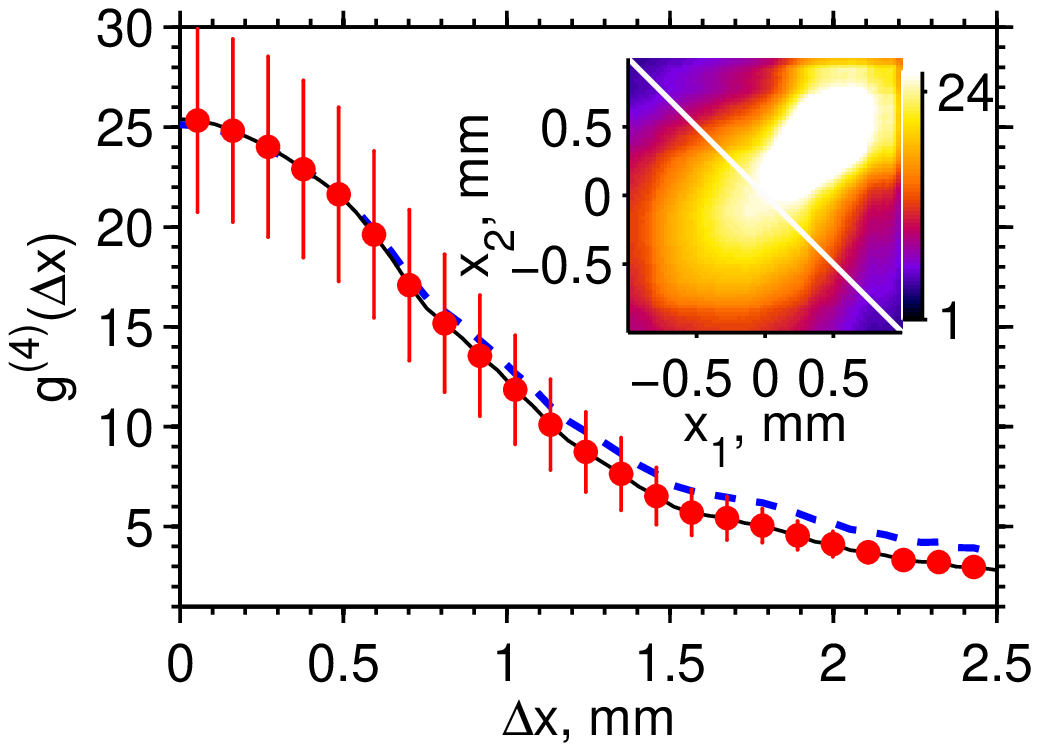}\\
  \includegraphics[height=0.15\textheight]{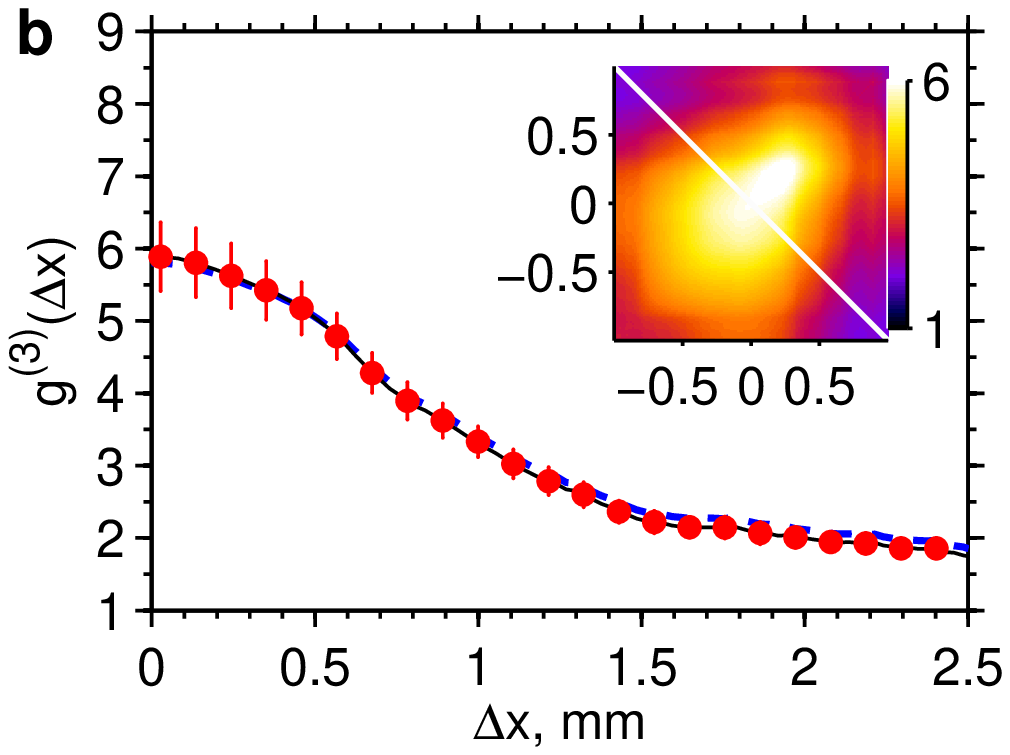}
  \includegraphics[height=0.15\textheight]{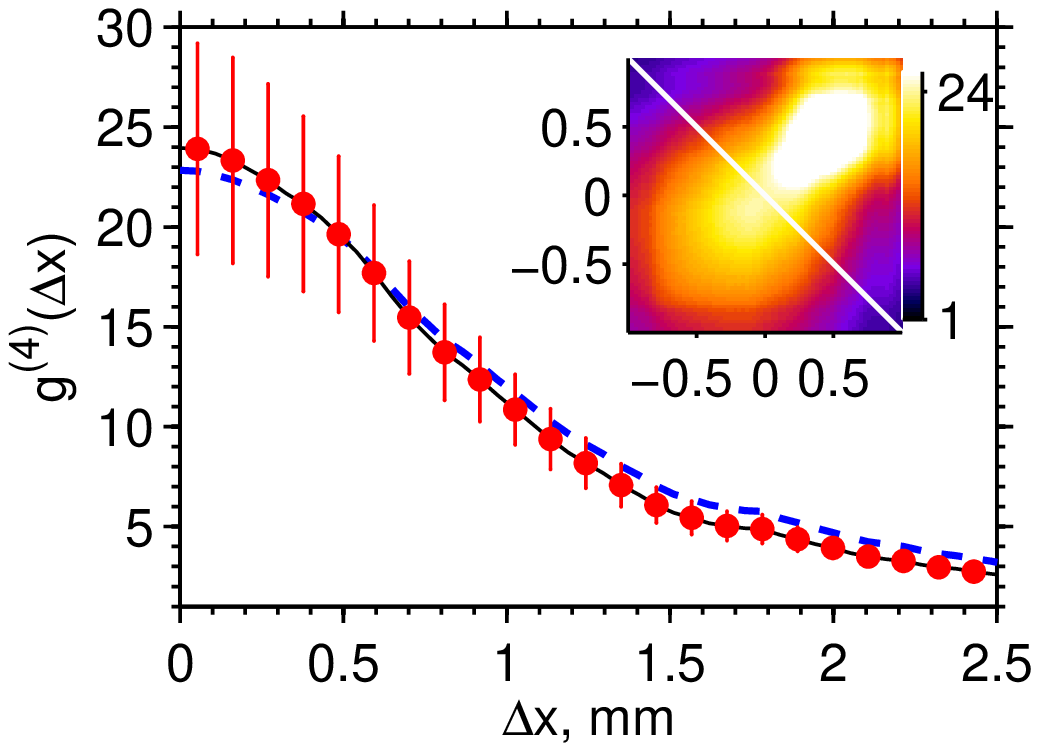}\\
  \includegraphics[height=0.15\textheight]{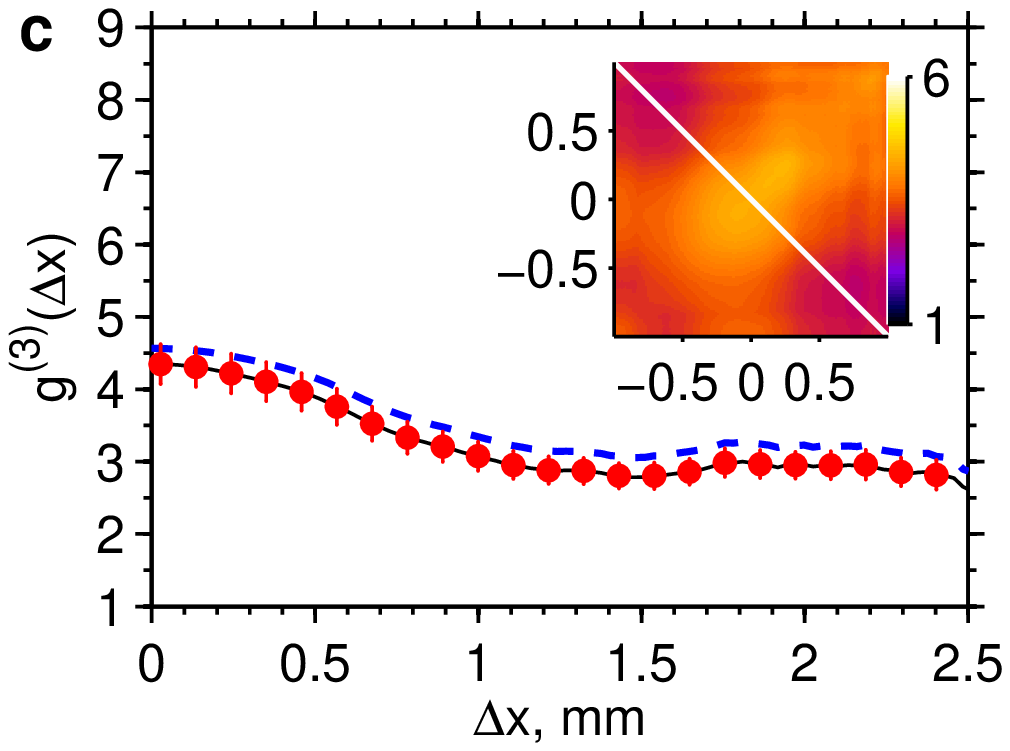}
  \includegraphics[height=0.15\textheight]{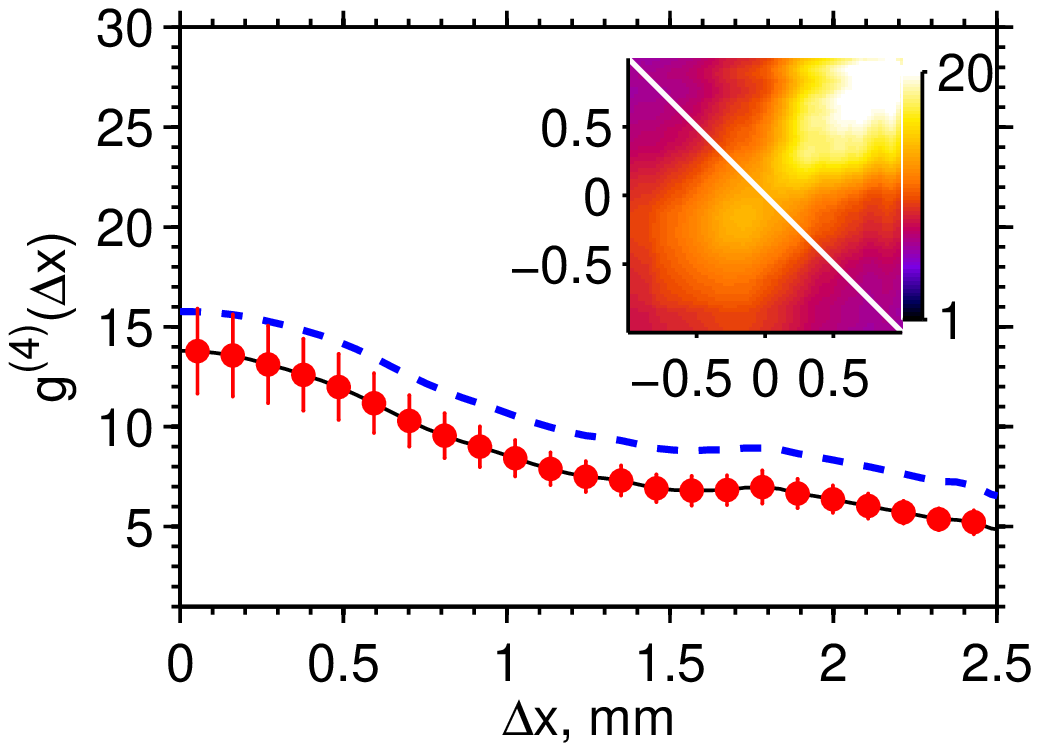}\\
  \includegraphics[height=0.15\textheight]{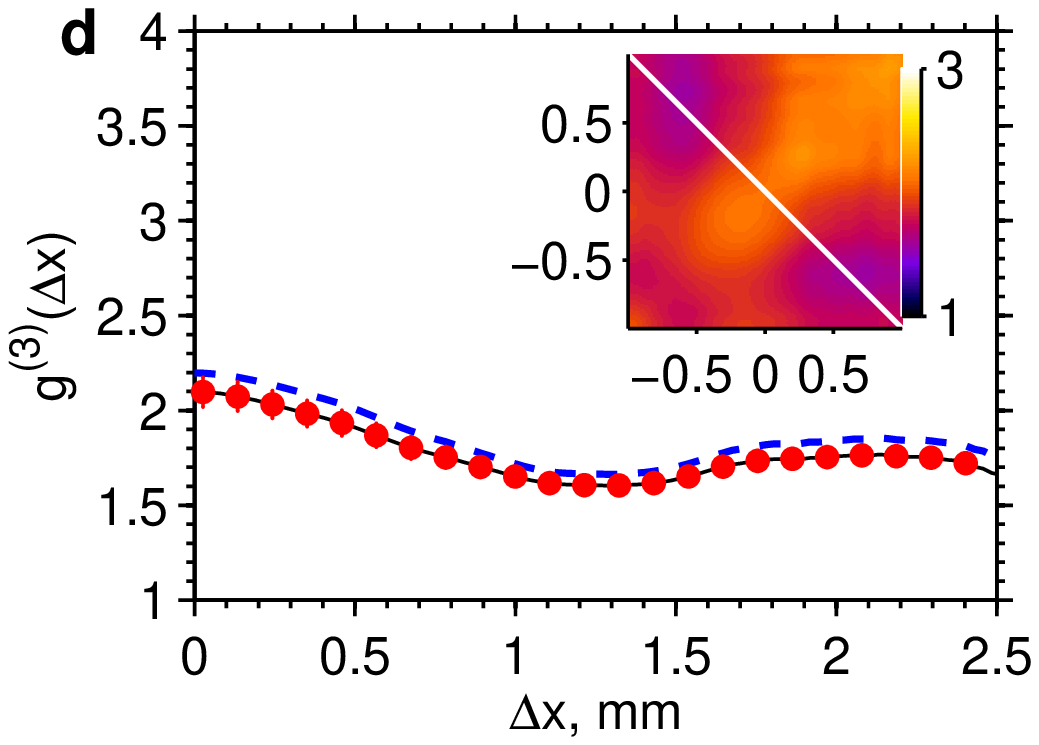}
  \includegraphics[height=0.15\textheight]{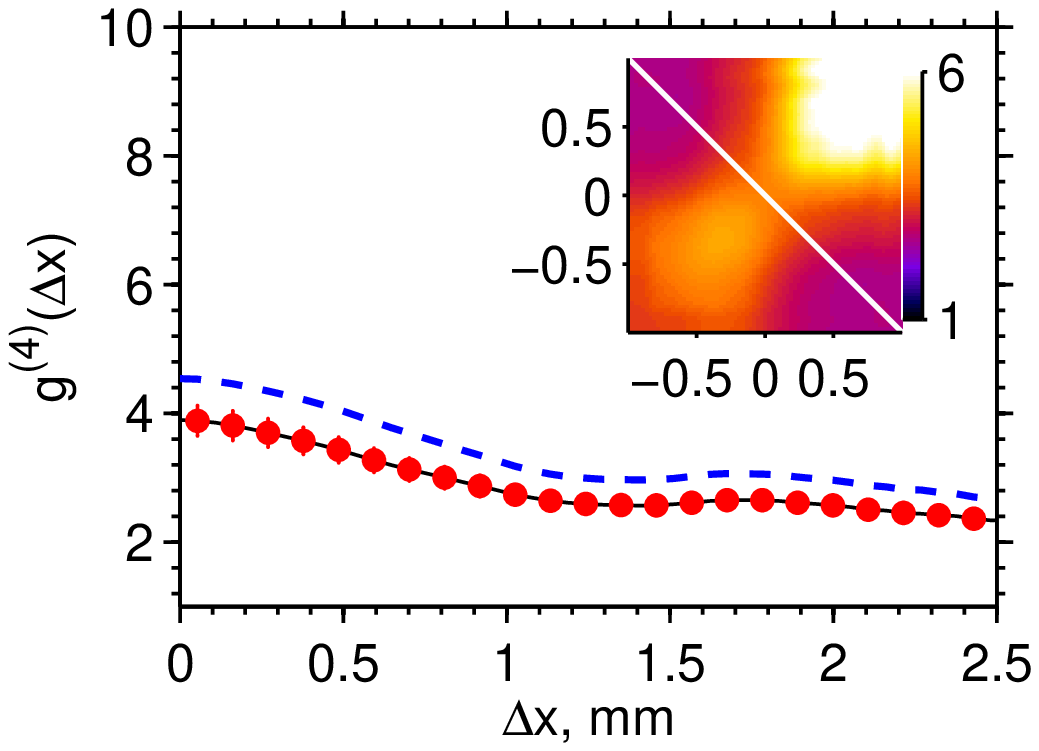}\\
  \includegraphics[height=0.15\textheight]{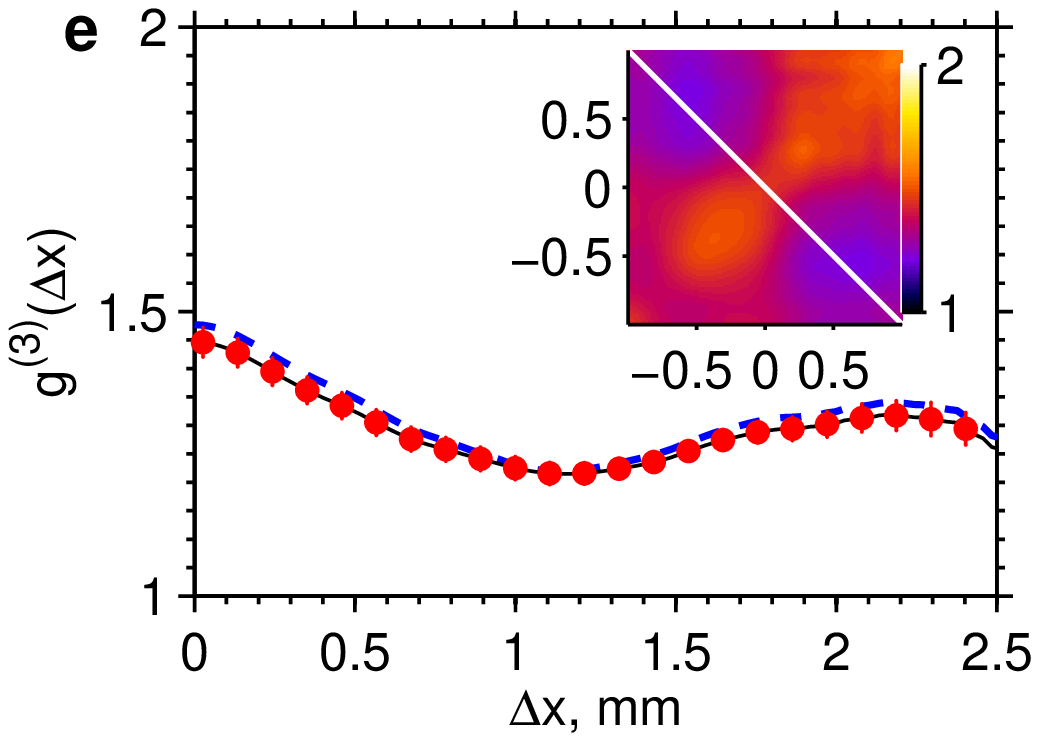}
  \includegraphics[height=0.15\textheight]{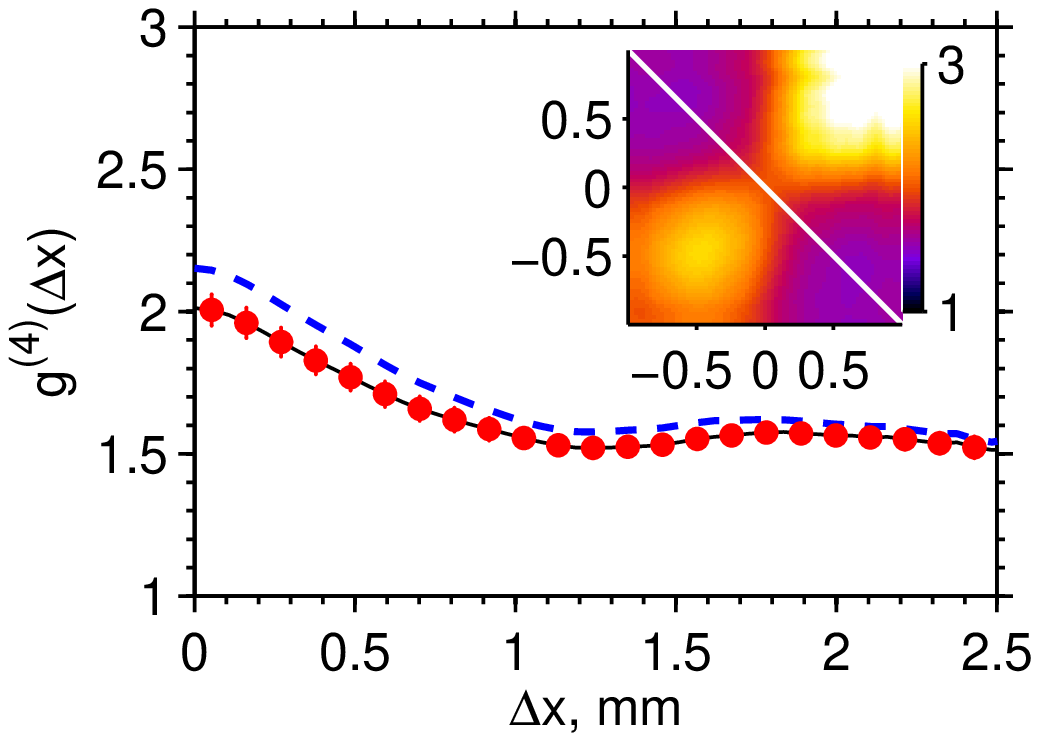}
  \caption{%
  Same as in Fig.~4 of the main text for all measured monochromator settings.
  The monochromator bandwidth $\Delta E/E$ was (a) $0.8\cdot10^{-4}$, (b) $1.7\cdot10^{-4}$, (c) $4\cdot10^{-4}$, (d) $8\cdot10^{-3}$, and (e) $1.4\cdot10^{-3}$.
  Third-order correlation functions (left column)
  fourth-order correlation functions (right column). In the insets two-dimensional distributions of the corresponding correlation functions are presented.
  }
  \label{fig:S2}
\end{figure}

We also present the second-order correlation function for all measured monochromator settings in Fig.~\ref{fig:S1}.
The histogram distributions of the total intensity per pulse are also shown in the same Figure.
Experimental results for the third and fourth-order intensity correlation functions as well as comparison with the expressions
\eqref{eq:correlation3_final} and \eqref{eq:correlation4_final} for different monochromator settings are presented in Fig.~\ref{fig:S2}.


\newpage



\begin{thebibliography}{99}
\bibitem{1}	R. Hanbury Brown and R. Q. Twiss, Nature \textbf{177}, 27-29 (1956).
\bibitem{2}	R. Hanbury Brown and R. Q. Twiss, Nature \textbf{178}, 1046-1048 (1956).
\bibitem{3}	R. J. Glauber, Phys. Rev. \textbf{130}, 2529–2539 (1963).
\bibitem{4}	F. T. Arecchi, E. Gatti, and A. Sona, Phys. Lett. \textbf{20}, 27-29 (1966).
\bibitem{5}	M. A\ss{}mann, F. Veit, M. Bayer, M. van der Poel, and J. M. Hvam, Science \textbf{325}, 297-300 (2009).
\bibitem{6}	S. S. Hodgman, R. G. Dall, A. G. Manning, K. G. H. Baldwin, and A. G.  Truscott, Science \textbf{331}, 1046-1049 (2011).
\bibitem{7}	A. Perrin, R. B\"ucker, S. Manz, T. Betz, C. Koller, T. Plisson, T. Schumm, and J. Schmiedmayer, Nature Physics \textbf{8}, 195-198 (2012).
\bibitem{8}	W. Ackermann \textit{et al.}, Nature Photon. \textbf{1}, 336-342 (2007).
\bibitem{9}	P. Emma \textit{et al.},  Nature Photon.  \textbf{4}, 641-647 (2010).
\bibitem{10}	T. Ishikawa \textit{et al.},  Nature Photon. \textbf{6}, 540-544 (2012).
\bibitem{11}	M. Altarelli \textit{et al.}, (eds.) The European X-Ray Free-Electron Laser Technical Design Report, DESY 2006-097, 2006, http://xfel.desy.de/technical\_information/tdr/tdr/.
\bibitem{12}	H. Chapman \textit{et al.}, Nature \textbf{470}, 73-77 (2011).
\bibitem{12a} K. J. Gaffney, H. N. Chapman, Science \textbf{316}, 1444-1448 (2007).
\bibitem{13}	M. Seibert \textit{et al.}, Nature \textbf{470}, 78-81 (2011).
\bibitem{14}	L. Mandel and E. Wolf, \textit{Optical Coherence and Quantum Optics} (Cambridge University Press, Cambridge, England, 1995).
\bibitem{14a}	J. W. Goodman, \textit{Statistical Optics} (Wiley, New York, 1985)
\bibitem{15}	A. Singer, I. Vartanyants, M.Kuhlmann, S. Duesterer, R. Treusch, and J. Feldhaus,  Phys. Rev. Lett. \textbf{101}, 254801 (2008).
\bibitem{16}	R. Mitzner \textit{et al.}, Opt. Express \textbf{16}, 19909-19919 (2008).
\bibitem{17}	W. F. Schlotter, F. Sorgenfrei, T. Beeck, M. Beye, S. Gieschen, H. Meyer, M. Nagasono, A. F\"ohlisch, and W. Wurth, Opt. Lett. \textbf{35}, 372-374 (2010).
\bibitem{18}	I. A.Vartanyants \textit{et al.}, Phys. Rev. Lett. \textbf{107}, 144801 (2011).
\bibitem{19}	A. Singer \textit{et al.}, Opt. Express \textbf{20}, 17480-17495 (2012).
\bibitem{20}	C. Gutt \textit{et al.}, Phys. Rev. Lett. \textbf{108}, 024801 (2012).
\bibitem{21}	H. Paul, Rev. Mod. Phys. \textbf{58}, 209-231 (1986).
\bibitem{23}	E. Ikonen, Phys. Rev. Lett. \textbf{68}, 2759-2761 (1992).
\bibitem{24}	E. L. Saldin, E. A. Schneidmiller, and M. V. Yurkov, \textit{The Physics of Free Electron Lasers} (Springer-Verlag, Berlin, 2000).
\bibitem{35}	M. Martins, M. Wellh\"ofer, J. T. Hoeft, W. Wurth, J. Feldhaus, and R. Follath, Rev. Sci. Instrum. \textbf{77}, 115108 (2006).
\bibitem{36}	N. Gerasimova, S. Dziarzhytski, and  J. Feldhaus, J. Mod. Opt. \textbf{58}, 1480-1485 (2011).
\bibitem{25}	E. Gluskin, E. E. Alp, I. McNulty, W. Sturhahn, and J. Sutter, J. Synchrotron Rad. \textbf{6}, 1065-1066 (1999).
\bibitem{26}	M. Yabashi, K. Tamasaku, and T. Ishikawa, Phys. Rev. Lett. \textbf{87}, 140801 (2001).
\bibitem{27}	M. Yabashi, K. Tamasaku, and T. Ishikawa, Phys. Rev. Lett. \textbf{88},  244801 (2002).
\bibitem{37}	E. L. Saldin, E. A., Schneidmiller, and M. V. Yurkov, Opt. Commun. \textbf{281}, 1179-1188 (2008).
\bibitem{38}	I. A. Vartanyants and A. Singer, New J. Phys. \textbf{12}, 035004 (2010).
\bibitem{28}	G. Baym, Acta Physica Polonica B \textbf{29}, 1839-1883 (1998).
\bibitem{29}	A. A. Lutman, Y. Ding, Y. Feng, Z. Huang, M. Messerschmidt, J. Wu, and J. Krzywinski, Phys. Rev. STAB \textbf{15}, 030705 (2012).
\bibitem{30}	Y. Inubushi \textit{et al.},  Phys. Rev. Lett. \textbf{109}, 144801 (2012).
\bibitem{31}	U. Fr\"uhling \textit{et al.}, Nature Photon. \textbf{3}, 523-528 (2009).
\bibitem{32}	G. Geloni, V.  Kocharyan, and E. Saldin, J. Mod. Opt. \textbf{58}, 1391-1403 (2011).
\bibitem{33}	J. Amann \textit{et al.}, Nature Photon. \textbf{6}, 693-698 (2012).
\bibitem{34}	R. Neutze, R. Wouts, D. van der Spoel, E. Weckert, and J. Hajdu,  Nature \textbf{406}, 752-757 (2000).
\bibitem{LajunenJOSA2005} H. Lajunen, P. Vahimaa, and Jani Tervo, J. Opt. Soc. Am. A, \textbf{22}, 1536 (2005)

\end{thebibliography}
\end{document}